\documentclass[preprint]{aastex}

\newcommand{\Lya}{\hbox{Ly$\alpha$}}

\newcommand{\Elo}{\hbox{$E^{\prime\prime}$}}
\newcommand{\kms}{\hbox{km s$^{-1}$}}
\newcommand{\erg}{\hbox{erg cm$^{-2}$ s$^{-1}$}}
\newcommand{\erga}{\hbox{erg cm$^{-2}$ s$^{-1}$ \AA$^{-1}$}}
\newcommand{\IUE}{\textit{IUE}}

\newcommand{\HST}{\textit{HST}}
\newcommand{\STIS}{STIS}            % supposed to be roman typeface -JAV
\newcommand{\FUSE}{\textit{FUSE}}
\slugcomment{submitted to AJ}
\shorttitle{Loopy UV line profiles of RU Lupi}
\shortauthors{Herczeg et al.}

\begin{document}

% Title.
\title{The loopy UV line profiles of RU Lupi: accretion,
outflows, and fluorescence}

% Authors.
\author{Gregory J. Herczeg\altaffilmark{1}, Frederick M. Walter\altaffilmark{2}, Jeffrey L. Linsky\altaffilmark{1}, G\"osta F. Gahm\altaffilmark{3}, David R. Ardila\altaffilmark{4}, Alexander Brown\altaffilmark{5}, Christopher M. Johns-Krull\altaffilmark{6}, Michal Simon\altaffilmark{2}, Jeff A. Valenti\altaffilmark{7}}

\altaffiltext{1}{JILA,  University of Colorado and NIST, Boulder, CO 80309-0440;  gregoryh@casa.colorado.edu, jlinsky@jila.colorado.edu}

\altaffiltext{2}{Department of Physics and Astronomy, Stony Brook University, Stony
Brook NY 11794-3800; fwalter@astro.sunysb.edu, msimon@astro.sunysb.edu}

\altaffiltext{3}{Stockholm Observatory, AlbaNova, SE - 106 91 Stockholm, Sweden; gahm@astro.su.se}

\altaffiltext{4}{Department of Physics \& Astronomy, Johns Hopkins University,
Baltimore MD 21218; ardila@pha.jhu.edu}

\altaffiltext{5}{Center for Astrophysics and Space Astronomy, University of Colorado, 
Boulder CO 80309-0389; ab@casa.colorado.edu}

\altaffiltext{6}{Department of Physics \& Astronomy, Rice University, Houston TX
77005-1892; cmj@rice.edu}

\altaffiltext{7}{Space Telescope Science Institute, Baltimore MD 21218; valenti@stsci.edu}

% Abstract
\begin{abstract}
We present far-ultraviolet spectra of the classical T Tauri star RU Lupi
covering the 912--1710 \AA\ spectral range,
as observed by the \HST/STIS and \FUSE\ satellites.  We use these spectra, which are rich in emission and
absorption lines, to probe both the accreting and outflowing gas.
Absorption in the \Lya\ profile constrains the extinction to $A_V\sim0.07$
mag, which we confirm with other diagnostics.  We estimate a mass
accretion rate of $(5\pm2)\times10^{-8}$ $M_\odot$ yr$^{-1}$ using the optical-NUV accretion continuum. 
The accreting gas is also detected in bright, broad lines of 
 \ion{C}{4}, \ion{Si}{4}, and \ion{N}{5}, which all show complex structures across the line profile.  Many other emission lines,
including those of H$_2$ and \ion{Fe}{2}, are pumped by \Lya.  RU Lupi's
spectrum varies significantly in the FUV; our \STIS\ observations
occurred when RU Lupi was brighter than several other observations in the FUV, possibly due to a high mass
accretion rate.
\end{abstract}

% Official ApJ keywords in alphabetical order.
% See http://www.journals.uchicago.edu/ApJ/information.html

% Citation commands:
%
% \citet{Nam98} for text citations, e.g. "Name (1998) says..."
% \citep{Nam98} for parenthetical citations, e.g. "is true (Name 1998)."
% \citep{Nam98,Alt01} to cite multiple refs, e.g. "(Name 1998; Alt 2001)"
% \citep{Nam98,Nam01} automatically merges, e.g. "(Name 1998; 2001)"
% \citet*{Nam98} or \citep* for the first citation of 3 author papers.
% \citet*{Nam98} or \citep* for the first citation of 3 author papers.
% \citep[e.g.,][]{Nam98} for leading text, e.g. "(e.g., Name 1998)"
% \citep[, for example]{Nam98} for trailing text.
% \citep[e.g.,][, for example]{Nam98} for leading and trailing text.
% \citealt{Nam98} is \citet{Nam98} without parentheses.
% \citealp{Nam98} is \citep{Nam98} without parentheses.
% \citeauthor{Nam98} only lists authors, e.g. "Name"
% \citeyear{Nam98} only lists the year, e.g. "1998"
% \citeyearpar{Nam98} only lists the year in parentheses, e.g. "(1998)"
% \citetext{\citealp{Nam98}; see also \citealp{Alt01}} embeds text, e.g.
%         "(Name 1998; see also Alt 2001)"
%
% Look up "natbib" on the web for a more detailed description.

%%%%%%%%%%%%%%%%%%%%%%%%%%%%%%%%%%%%%%%%%%%%%%%%%%%%%%%%%%%%%%

\section{INTRODUCTION}
Classical T Tauri stars (CTTSs) are likely prototypes of the young Sun when it was accreting
material from its circumstellar disk.  In the standard magnetospheric
accretion model, the magnetic field truncates the disk at a few stellar
radii.  Material from the disk accretes onto the star along the magnetic field lines and
is shocked as it decelerates at the stellar surface.  A fast outflow is
likely associated with the accretion process, and a slow outflow probably
emanates from the disk \citep[see discussion in, e.g., ][]{Har95}.

Far-ultraviolet (FUV) spectra reveal these diverse processes.  
\citet*{Joh00} and \citet{Cal04} found a correlation between the mass accretion rate and the strength of
\ion{C}{4} emission from CTTSs.
\ion{C}{4} and other high temperature emission lines are much weaker
in non-accreting naked T Tauri stars (NTTSs) than in CTTSs \citep{Joh00,Ard02a}.  Thus,
these lines, which trace gas at $10^5$ K, are likely produced by the
accretion shocks.

\citet{Ard02b} detected wind absorption in the
profiles of \ion{Mg}{2} lines in \HST/GHRS spectra of 8 CTTSs, sampling a range of inclinations.  They
found that the equivalent width of the wind absorption is larger for stars
with disks that are observed face-on than for stars with disks observed edge-on, suggesting some collimation in the outflow.  Similar P-Cygni profiles
of strong lines of \ion{O}{1} and \ion{C}{2} are typically
found in FUV spectra of CTTSs.  Fluorescent
H$_2$ emission may trace both warm molecular gas at or near the disk surface
\citep{Her02} and molecular gas in the surrounding cloud that is shocked as 
it interacts with stellar outflows \citep{Wal03,Sau03}.

In this paper, we present \HST/\STIS\ and \FUSE\ spectra, covering the
912--1700 \AA\ spectral range, of RU Lupi, one of the most heavily veiled$^1$  CTTSs at optical wavelengths.  The
spectrum is rich in strong absorption features and in emission lines
produced by hot and cool gas.  In \S 2, we describe the properties of RU Lupi.  In \S 3, we describe our {\it HST}/STIS and {\it FUSE} observations.  In \S 4, we measure the total hydrogen column density in our line
of sight to RU Lupi to estimate the extinction.  In \S 5, we measure
the mass accretion rate and present hot emission in lines that are likely
produced by the accreting gas.  In \S 6, we show wind absorption in line profiles.  Finally, in \S 7, we
analyze variability in the FUV emission from RU Lupi.  Our conclusions
   are presented in \S 8.
\footnotetext[1]{ratio of accretion flux to
photospheric flux, see \citet{Joy49,Her62}}

\section{Properties of RU Lupi}
RU Lupi (HD142560) has some of the strongest accretion signatures among CTTSs, including strong optical veiling and an H$\alpha$ 
equivalent width that has been observed to vary in the range 140--216 \AA\
\citep[e.g.][]{App83,Lam96}.
\citet{Ste02} estimate an optical veiling
of 10--20 at 4091 \AA\ 
and 2--5 between 5300--6300 \AA.
This strong veiling, the result of a high accretion rate, makes it
difficult to estimate stellar properties from the uncertain photospheric emission.
\citet{Ste02} extracted the optical photospheric spectrum of RU Lupi from
high-resolution {\it VLT}/UVES spectra, and measured a radial velocity of
$-1.9\pm1$ \kms.  They estimate
$T_{eff}=3950$ K and $\log g=3.9$.  Using these numbers, and an approximate 
mass of  $M=0.8$ $M_\odot$ \citep{Lam96}, they
calculate $R=1.66$ $R_\odot$ and $L=0.49$
$L_\odot$, assuming a distance of 200 pc.  The distance is probably closer
to 140 pc \citep{dez99,Ber99}.  Based on our extinction estimate $A_V=0.07$ 
(see \S 4), $d=140$ pc, and the temperature and V-band veiling estimated by
\citet{Ste02}, we calculate a photospheric luminosity $L_{phot}=0.6$
$L_\odot$ and $R=1.64$ $R_\odot$.  Given these characteristics,
the \citet{Sie00} pre-main sequence stellar evolution tracks indicate that RU
Lupi has a mass of $0.6-0.7$ M$_\odot$ and an age of 2--3 Myr.

Although most previous studies have assumed that RU Lupi is a single star,
\citet*{Gah04} used the {\it VLT} to
detect radial velocity changes of 2.5 \kms\ in photospheric
absorption lines that are consistent with a period of
either 3.7 or 5.6 days.  They suggest that these changes are not due to
spots, and may instead indicate the presence of a close stellar companion.  RU Lupi 
is variable at ultraviolet and optical wavelengths.  A
photometric period of 3.7 d was also found by 
\citet{Hof58,Hof65}, but no period was confirmed in long-term monitoring
\citep[see][]{Gio95}.
\citet{Hut89} and \citet{Gio90} found 
that RU Lupi is redder when it is fainter, which suggests that the
brightening could be due to either an increase in the mass accretion rate, 
a decrease in circumstellar extinction, or cool starspots.

\citet{Ste02} used optical metallic absorption lines
to measure a rotational velocity $v \sin i=9\pm1$ \kms.
Rotational velocities of CTTSs can be at least $v\sin i=65$~\kms\ \citep{Har89,Bas90,Joh01}.
The measured rotational velocity suggests that either the rotational velocity is small or the rotation axis, and presumably the disk, are observed close to face-on.
\citet{Ard02a} found that the
equivalent width of the \ion{Mg}{2} wind absorption in NUV spectra of RU Lupi is the largest of the 8
CTTSs in their sample, which we would expect if the wind is collimated and the disk is observed face-on.
Blueshifted emission in profiles of the optical [\ion{O}{1}], \ion{Na}{1}, and
\ion{Ca}{2} H and K lines suggest that the disk obscures wind emission on the far side of the star, supporting
the face-on viewing angle to the star \citep{Ste02}.  Based on the observations described above, we infer that the disk is probably observed close to face-on.  If the 3.7 d period is due to rotation and not a close compation, the rotational broadening of absorption lines would imply an inclination of 24$^\circ$.
Spectro-astrometry of RU Lupi shows that H$\alpha$ emission may be streaming SW and NE from
the star \citep{Tak01}.  HH55, located 2$^\prime$ SW of the star, may be
related to outflows from RU Lupi, which would imply that the disk is at
least slightly inclined to our line of sight.  The temperature and density
of the base of the outflows from RU Lupi were also probed using {\it
HST}/STIS observations of emission in \ion{Si}{3}] 1892 \AA\ \citep{Gom01}.

The {\it ROSAT} count rate of RU Lupi was $\sim 9\times10^{-3}$ cts s$^{-1}$ in a 15.8 ks PSPC observation (0.1--2.4 keV).  We convert this to an X-ray flux of $8\times10^{-14}$ \erg, or $L_X=5\times10^{-5} L_\odot$ using \citet{Neu95}.

RU Lupi is associated with a molecular cloud, mapped in $^{12}$CO
($J=2-1$) and $^{13}$CO ($J=1-0$) by \citet{Gah93} and in  $^{12}$CO ($J=2-1$ by \citet{Tac01}.
\citet{Gah93} found no direct evidence of any molecular outflows from RU Lupi,
although they note a possible connection with HH 55.  Both studies found
that the radial velocity of the molecular gas surrounding RU Lupi is about
-0.4 \kms.

\section{OBSERVATIONS}
\subsection{HST/STIS} 
We observed RU Lupi with \HST/\STIS$^1$ as part of \HST\ program
8157 (P.I. F. Walter).  The spectra consist
 of five $\sim 2500$ s exposures with the E140M echelle
grating, covering 1150--1710 \AA, and a 120 s
optical exposure with the G430L grating.  We used the $0\farcs2\times 0\farcs 06$
aperture to minimize contamination by any off-source
emission.  The FUV observations use a
1024$\times$1024 pixel CsI
MAMA detector, and the optical observations use a CCD.  
The pixel size on the MAMA detector is $0\farcs036$ in the dispersion direction (which corresponds to about 0.0142 \AA\ or 3.18 km s$^{-1}$ at 1330
\AA) and $0\farcs029$ in the cross-dispersion direction.  Details of these observations are listed in Table 1.
\footnotetext[2]{See \citet{Lei01} for a description of STIS.}

The STIS FUV spectra were obtained in time-tag mode.  Figure 1 shows that
the count rate is similar at the
beginning of each orbit but steadily increases within
each orbit by about 40\%.  
Figure 2 shows that the width of the point-spread function in the cross-dispersion 
direction, measured in regions of strong line emission, decreased during
these observations.  
We conclude that the telescope focus
improved during each orbit.  
Thermal variations of {\it HST} can change the telescope focus.  Models of
thermal focus variations$^3$ indicate that the focus was poor at the start
of each orbit but improved during every orbit.
The nominal FWHM of the point-spread function in the FUV-MAMA
is about 2.5 pixels.
Because the size of the aperture
($0\farcs2\times0\farcs06$) in the dispersion direction is smaller than the
line-spread function, the count rate increases as the focus improves.
\footnotetext[2]{See
http://www-int.stsci.edu/instruments/observatory/focus/focus2.html\#breathing 
for a description of modeling the HST focal length variations.}

Several other observations obtained using the E140M grating and the
$0\farcs2\times0\farcs06$ aperture on STIS, including those of V471
Tau, a white dwarf with a K2V companion \citep*{Obr01}, and several CTTSs also suffer from this problem.
On the other hand, observations of the two subluminous B stars PG1144+615 and
PG1219+534 record a constant count rate during each orbit.  No other STIS observations were taken in
time-tag mode with the $0\farcs2\times0\farcs06$ aperture. 
This effect may also be seen data taken with small apertures that are not in time-tag mode.

V471 Tau was observed at three
different epochs, twice with the $0\farcs2\times0\farcs06$ aperture and
once with the $0\farcs2\times0\farcs2$ aperture \citep[see Table 2,][]{Obr01,Wal04}.  The white
dwarf dominates FUV emission and has a 555 s period \citep{Jen86}, during
which its FUV flux varies by a few percent \citep{Wal04}.  V471 Tau is also
a bright FUV source,
making it excellent to use for calibration.  
Figure 3 shows the total flux from 1250--1600 \AA\ in 300 s
intervals of four orbits from the observations of V471 Tau.  We find one
exposure taken with the small aperture, STIS observation number 24906,
during which the detected flux is constant, at $4.01\times 10^{-10}$ erg
cm$^{-2}$ s$^{-1}$.  Certain exposures, such as observation numbers 97314
and 97316, show flux increases at a rate about half that found in the
observations of RU Lupi.
These observations have a peak flux of $3.92\times 10^{-10}$ erg cm$^{-2}$
s$^{-1}$.  The observations taken with the large aperture have a flux
1250--1600 \AA\ of $4.46\times10^{-10}$ erg cm$^{-2}$ s$^{-1}$. 

We measure the flux and cross-dispersion point-spread function in each 300 s
interval for the observations of RU Lupi and V471 Tau.  The point-spread
function depends on wavelength, so we measure the FWHM of emission in the
cross-dispersion direction at wavelengths with strong emission lines (\ion{O}{1}, \ion{C}{2},
\ion{Si}{4}, and \ion{C}{4}) in the spectra of RU Lupi, and in the same
regions in the spectra of V471 Tau.  Figure 4 shows the percent of the
total flux between 1230--1650 \AA\ as a function of point-spread function for V471 Tau and RU Lupi.
We calibrate the flux of RU Lupi by calculating a best-fit line to the data
from V471 Tau, and then minimizing square of the difference between the data 
from RU Lupi and the best-fit line calculated from the V471 Tau observations.

For RU Lupi, we calculate a total flux of $6.02\times10^{-12}$ erg
cm$^{-2}$ s$^{-1}$ between 1230-1650 \AA\ and  $8.43\times10^{-13}$ erg
cm$^{-2}$ s$^{-1}$ between 1545--1555 \AA\ (a
region dominated by \ion{C}{4} emission).
We estimate that this flux calibration
has an absolute uncertainty of about 15\%.
 Inspection of the V471 Tau observations indicate
that the changes in flux are constant with wavelength, such that the
relative flux calibration across the wavelength range is uncertain to $<3\%$
within a single exposure and between different exposures.  
The consistency of the relative count rate in strong lines during our
observation of RU Lupi indicates that the relative flux is uncertain by
$<10\%$.

The Doppler correction in our data was in error by a factor of 1.6, such
that the spectrum shifts by 3.8 pixels (12 \kms) during a single orbit.  We
corrected for this error by subdividing the spectra into 300 s intervals,
and subsequently shifting and coadding the spectra.  This error is
prevalent in STIS echelle spectra in the FUV, but is now
corrected in newer versions of the {\it calstis} pipeline.$^4$   
The spectral resolution is about $R=45,000$. 
\footnotetext[4]{see http://www.stsci.edu/hst/stis/calibration/pipe$\_$soft$\_$hist/update215c.html}

The wavelengths were initially calibrated using short exposures of the
on-board Pt/Cr-Ne lamp obtained at the end of each observation of RU Lupi.
We determined the absolute wavelength calibration independently by
measuring the wavelength of the geocoronal Ly$\alpha$ emission line.  We then shift the line to 1215.58 \AA, where we expect the
line to be at the time of our observation.  We find
that our wavelength solution corresponds well to that based on the
calibration lamp.  The absolute and relative wavelength solution is  accurate to at least
3 \kms.

\subsection{FUSE}
We obtained a 25 ks spectrum of RU Lupi with the \FUSE\ satellite in
21 separate exposures, each covering 912--1187 \AA, as part of \FUSE\
program A109 (P.I. F. Walter).  These observations
occurred at night to minimize contamination by airglow emission.  The {\it
FUSE} has four co-aligned telescopes.  The optical paths are coated 
with SiC or Al:LiF to maximize efficiency between 900--1000 \AA\ or
1000-1187 \AA, respectively. One observation produces eight
overlapping spectra, each of which covers about 90 \AA.  A detailed description of the \FUSE\ mission can be found in \citet{Moo00}
and \citet{Sah00}.
Since we used the
LWRS aperture ($30^{\prime\prime}\times30^{\prime\prime}$), the \FUSE\
spectra would include emission extended beyond the point source of RU Lupi, if present.

The exposures were coadded and reduced with the {\it calFUSE} v2.4 software 
package \citep*{Kru01}.
 The 8 separate spectra were then coadded using {\it hrs\_merge} in {\it
IDL}.  No significant differences were found in overlapping wavelength
ranges of these segments.
Since the LiF channels are more sensitive than
the SiC channels, the S/N in the \FUSE\ spectrum is higher at $\lambda>1000$ \AA.
The data were binned by 4 pixels, which is smaller than the approximate
instrument resolution of 9 pixels, or $R\approx15,000$.  
The flux scale is
accurate to about $\pm$5\%, including a small degradation in some of the
channels over time.

We calibrate wavelengths in our \FUSE\ spectrum by comparing the measured
wavelengths of H$_2$ emission lines (see \S 5.2) in the \FUSE\ spectrum to those in the STIS spectrum.  The H$_2$ lines in the
STIS spectrum can be characterized by a bright narrow component and a fainter and broader blueshifted component.  When
convolved to the lower spectral resolution of \FUSE, the \STIS\ H$_2$ lines appear Gaussian with a
velocity shift of -15.5 \kms\ relative to the photospheric radial velocity of RU Lupi.
We offset the pipeline-calibrated wavelengths by +5 \kms, so that the
average H$_2$ emission line in the \FUSE\ spectrum has a velocity shift of -15.5
\kms.  
This velocity shift also places the peak of the \ion{N}{1} 1135~\AA\
emission line at the radial velocity of the star, which is similar to the velocity shift of emission lines from neutral atoms in the \STIS\ spectrum (see \S 5.2).   
The shift of 5 \kms\ is less than the 15 \kms\ accuracy of the standard
{\it calFUSE} pipeline wavelength calibration \citep*{Kru01}.
We estimate an uncertainty of about 5 \kms\ in our wavelength calibration at $\lambda>1100$ \AA, and a larger uncertainty at $\lambda<1100$ \AA.

The \STIS\ spectrum overlaps the \FUSE\ spectrum between 1160--1185 \AA,
although the sensitivity of \STIS\ is poor at these short
wavelengths.  Figure 5 shows that the \ion{C}{3} line at 1175 \AA\ is much
stronger in the \STIS\ spectrum than in the \FUSE\ spectrum, so fluxes from the two spectra are not directly comparable.

\section{THE ISM IN OUR LINE OF SIGHT TO RU LUPI}
\subsection{Interstellar Absorption Lines}
Narrow absorption lines, produced in the interstellar medium (ISM), are detected
against several stellar emission lines (see Table 3 and Figure 6).  In \S 5.2, we
find that most low-ionization emission lines have Gaussian profiles centered about the radial velocity of RU Lupi.  We therefore
assume that the intrinsic emission profile is Gaussian to measure the velocity and
equivalent width of the absorption.

Interstellar absorption is centered between $-8$ and $-19$ \kms\ in the heliocentric frame, depending on the line. 
RU Lupi is located at $l=338^\circ, b=+11^\circ89$, where models of the
local interstellar medium produced by \citet{Red00} predict that very
little absorption by the LIC is expected and some absorption by the G cloud
is possible.  These models also predict that any absorption in the local
interstellar cloud and the G cloud should be centered at -23 \kms\ and -26
\kms, respectively.  The Lupus molecular cloud has a velocity of about -0.4
\kms\ in our line of sight toward RU Lupi \citep{Gah93,Tac01}.
\citet{Tac01} estimate that the H$_2$ column density of the Lupus cloud in
the direction of RU Lupi is $\log N($H$_2)=21.0$, which is much higher than
the column density $19.0<\log N($H$_2)<19.5$ in our line of sight to RU
Lupi, suggesting that the star is located in front of most of the cloud.
The data show absorption components
redshifted relative these clouds, indicating that other interstellar clouds are located along the RU Lupi line of sight.

\citet{Ard02b} found that the equivalent widths of interstellar \ion{Mg}{2} 
lines towards CTTSs are larger than typical values measured through the ISM 
toward nearby sources.  We also detect large equivalent widths in many narrow absorption lines in the
FUV, including a detection of interstellar absorption in the \ion{Si}{3}
1206.5 \AA\ line.  Interstellar \ion{Si}{3} absorption has only been detected in
one other local sightline, towards $\epsilon$ CMa \citep{Gry95, Woo02b}.  The large \ion{Mg}{2} equivalent widths and detection of \ion{Si}{3} absorption may be produced by warm or shocked gas in the Lupus molecular cloud.   
Any additional absorption
components due to the Lupus molecular cloud, a remnant envelope, or a slow
wind should be centered within a few \kms\ of the interstellar absorption. 

Ly$\alpha$ is detected in emission only longward of 1217.5 \AA, 
indicating that the sum of interstellar and
circumstellar \ion{H}{1} absorption is optically
thick shortward of 1217.5 \AA.  
We do not detect any 
blueshifted emission from \Lya\ because of strong \ion{H}{1} absorption in the wind.  ISM absorption should dominate absorption on the red side of the line.
We model the observed Ly$\alpha$ profile with a single Gaussian emission
profile, centered at the radial velocity of the star, and a Voigt absorption profile.  The height of the Gaussian is determined from the best fit to the red side of the \Lya\ line for a given $N($\ion{H}{1}).  We
assume that the absorption is centered 
at -15 \kms\ in the heliocentric frame, roughly the velocity at which we
detect interstellar absorption in lines such as \ion{O}{1} and \ion{C}{2}
(Table 3).  We assume a Doppler broadening parameter $b\sim11$ \kms, assuming a temperature of $7000$ K, typical of the local ISM \citep{Red00}.  At the high column densities applicable to the RU Lupi sightline, the column density is insensitive to $b$.
Figure 7 compares the profiles that best fit the observed emission for a
range of neutral hydrogen column density $N$(\ion{H}{1}).
We find that $\log
N$(\ion{H}{1})$=20.0\pm0.15$ (cm$^{-2}$ units), where the error includes an estimate of the uncertainty in
the central velocity of the ISM absorption.  

The low S/N in the continuum prevents us from detecting most of the H$_2$
absorption lines that should occur in the \FUSE\ bandpass.
However, the 6-0
P(3) line at 1031.19 \AA\ is detected because it occurs against
strong \ion{O}{6} emission (Fig. 8).  This H$_2$ line is shifted by 
about $-23\pm5$ \kms\ relative to the stellar radial velocity.
The \ion{O}{6} 1037 \AA\ line is
weak and narrow because most of the line is absorbed by H$_2$ and
\ion{C}{2} (see Fig. 8).  The detection of the one H$_2$ line and only weak
\ion{O}{6} 1037 \AA\ emission imply that
$19.0<\log N($H$_2)<19.5$ for temperatures $100$ K$<T$(H$_2)<300$ K, which is
typical of interstellar molecular hydrogen \citep{Rac02}.  We therefore
measure a total hydrogen column density of $\log
N$(H)=$\log [N$(\ion{H}{1}$)+2\times N($H$_2)]=20.1\pm0.2$.  \citet{Ste02} measured optically thin
absorption in the \ion{Na}{1} line at 5890 \AA, and found $\log
N$(\ion{Na}{1})$=11.0$.  They used interstellar abundances from
\citet{Wel01} to estimate $\log N$(H)$\sim19.8$, which is approximately
what we measure.

\subsection{Dust Extinction}
Strong optical veiling due to accretion prevents a reliable estimate of the
extinction using photospheric emission from RU Lupi.  Instead, we use the
interstellar absorption discussed above to constrain the extinction.
Assuming a typical interstellar gas/dust ratio and dust grains described by
$R_V=3.09$ \citep*{Car89}, we convert the hydrogen column density of $\log N$(H)$=20.1\pm0.2$ to an
extinction $A_V\sim0.07$ mag using the relation described by \citet{Boh78}.  
This low extinction implies that RU Lupi is near the front edge of the
Lupus molecular cloud.
We adopt $A_V=0.07$ mag for the remainder of this paper.
We confirm a low extinction to RU Lupi using two other probes in our data.

The NUV and blue continuum of RU Lupi is produced by accretion
\citep[e.g.][]{Har91}.  We use the shape of the accretion continuum to constrain the extinction by modelling the continuum as emission from an isothermal slab, described in \S 5.1.
Model continua that best fit the observed UV-optical ($1650$ \AA $<\lambda<5700$ \AA) continuum are calculated for each assumed value of the extinction.
Figure 9 shows the estimated accretion
continuum, calculated independently for three assumed extinctions compared with the extinction-corrected NUV spectrum of RU Lupi.  When we use
$A_V>0.4$ mag, our models of the accretion continuum underestimate the flux at 2175 \AA\ because the
extinction bump at 2175 \AA\ increases the assumed intrinsic flux at that wavelength.

If instead most of the dust grains in our line of sight to RU Lupi are in
the Lupus molecular cloud, then the dust may differ from typical
interstellar dust.  The grain size should not differ significantly from grains typical of the local ISM.  Although  sightlines through the densest regions of molecular clouds can have $R_V=5.5$, \citet*{Ken94} and \citet{Whittet01} find that the sightlines through the Taurus and Ophiucus molecular clouds with $A_V<3.0$ do not vary significantly from the average interstellar value of $R_V=3.09$ \citep{Car89}.
However, the extinction law towards the peculiar B star HD29647 ($A_V=3.6$ mag), which lies behind Taurus, does not show the strong 
2175 \AA\ bump
\citep{Whi04} that is typically associated with small carbonaceous dust grains
\citep[e.g.][]{Dra93}.  The grains in the line of sight to HD29647 have an
$R_V=3.63$, which suggests that they are large relative to typical interstellar grains, described by $R_V=3.09$, \citep{Car89}.  
\citet{Cal04} applied the HD29647 extinction law to
analyze ultraviolet spectra of intermediate-mass T Tauri stars in Taurus.
If the extinction law of HD29647 applies to low-extinction targets such as RU Lupi, then the shape of the accretion continuum is not a good probe of extinction because of the absence of the 2175 \AA\ bump.

We also place an upper limit on the extinction by comparing the relative strength of
fluorescent H$_2$ emission lines (see \S 5.2), which occur between
1100--1650 \AA, to the same emission lines from TW Hya.  
Extinction is negligible to TW Hya, and both
stars are observed nearly face-on, so any geometrical effects on the H$_2$ fluxes
should be similar. In \S 5.2, we identify 90 unblended H$_2$ lines in the
\STIS\ spectrum of RU Lupi from 15 different upper levels.  In the 
\STIS\ spectrum of TW Hya, \citet{Her02} detected 146 H$_2$ lines from 19
different upper levels.  By comparing the relative flux in lines from the
same upper level for the two stars, we place an upper limit of $A_V<0.6$ mag toward RU Lupi for either a
typical interstellar extinction law or the interstellar extinction law of
HD29647.

Our estimate of a low extinction towards RU Lupi agrees with most previous estimates.  \citet{Gah79}
estimated $0.3<A_V<1.0$ mag using the {\it IUE} spectrum of RU Lupi.
\citet{Gio95} estimated an extinction of $A_V=0.3$ mag based on the 2200
\AA\ reddening bump in {\IUE} spectra, and
their result agrees with the extinction $A_V=0.4$ mag estimated by
\citet{Bro01} and \citet{Lam96}.  
Our extinction
estimate is much lower the extinction of $A_V=1.28$ mag calculated by
\citet{Hug94} based on optical colors, which are contaminated by emission
due to accretion.  

The molecular fraction of gas towards RU Lupi is between 17--40\%.  Most sightlines with a molecular fraction
$>0.01$ have $A_V>2.4$ mag \citep{Sav77,Rac02}.  However, this relationship may
not apply to environments around young stars.  A high molecular fraction
and low extinction were also found towards the Herbig Ae star HD104237
\citep{Gra04}.

\section{ACCRETION}

\subsection{Mass Accretion Rate}
We use the accretion luminosity, estimated from the UV excess and
the Balmer jump at 3700 \AA, to approximate the mass accretion
rate of RU Lupi \citep{Val93,Gul98}.  Our \HST\ observations consist of nearly
simultaneous optical and FUV spectra.  We obtain NUV spectra of
RU Lupi using a combination of archival observations from \HST/STIS, \HST/GHRS, and
{\IUE} (Table 4).  The NUV spectra are normalized by factors listed in
Table 4 to match the optical and FUV continua flux, ignoring regions of
strong emission lines.  We scale the optical STIS spectrum of V836 Tau, a
K7 WTTS, to use as a template for the photospheric emission from RU Lupi.
We estimate that $A_V=1.52$ mag towards V836 Tau, the average extinction
estimate from \citet{Ken95} and \citet{Whittet01}.

The methodology used to calculate the mass accretion rate is described by
\citet*{Val93} and \citet{Her04}. 
We model the spectrum of the accretion shock by approximating the emission 
from an isothermal, pure-hydrogen
slab (see Fig. 10), which fit the data well.  Other studies have also calculated the accretion continuum by modelling the accretion shock \citep{Cal98}.
The emission 
spectrum that best fits the observed spectra is produced by a slab with
temperature $T=9600$ K, electron density $n_e=3\times10^{14}$ cm$^{-3}$, 
slab surface area 
parameter (fraction of stellar surface area covered by accretion) $\delta=0.017$, thickness $l=2\times10^7$ cm, and a scaling
factor for the template photospheric spectrum of 0.39.  The slab has an
optical depth of 0.64 at 5000 \AA, which is less than the average optical
depth in slab models of blue emission from other CTTSs \citep{Bas89,Har91}.
This scaling factor 
corresponds to an optical veiling in the continuum of
a factor of 4--8 between 4000--4300 \AA, and about 2 at longer wavelengths.
We calculate an accretion
luminosity $L_{slab}=0.35$ L$_\odot$.
We then estimate a mass
accretion rate of
\begin{equation}
\dot{M_{acc}}=\frac{1.25 R_*L_{slab}}{GM_*}=(5\pm2) \times 10^{-8} \frac{d^2}{140^2} \frac{R}{1.65 R_\odot} \frac{0.8 M_\odot}{M} M_\odot yr^{-1}
\end{equation}
This mass accretion rate is large, but significantly smaller than previous
rates that were estimated assuming $A_V=1.3$ \citep[e.g.,][]{Hug94,Ard02a}.
Although mass accretion rates can be as high as $10^{-6}$ $M_\odot$ yr$^{-1}$ for
CTTSs, the mass accretion rate of $5\times10^{-8}$ $M_\odot$ yr$^{-1}$ is
high for CTTSs with ages of 2--3 Myr \citep{Har98}.

The model accretion continuum accounts for the observed optical and NUV continuum.  A slight FUV excess may be detected, although the model accretion continuum in the FUV is sensitive to uncertainties in the temperature of the accreting gas.  Figure 11 shows that the FUV continuum from RU Lupi, if present, is much weaker than that detected towards TW Hya and several other CTTSs
\citep{Her04,Ber04}.  The FUV continuum from CTTSs is most likely produced by H$_2$ fluorescence generated
by highly energetic electrons colliding with and electronically exciting
H$_2$ molecules, and may be related to an inner dust hole in the disk
\citep{Liu02,Ber04}.  A bump at 1600 \AA, diagnostic of this H$_2$
emission, is weakly present.

\subsection{FUV Emission Lines}

The total observed emission between 1170--1700 \AA\ from RU Lupi is $7.16\times10^{-12}$ erg cm$^{-2}$ s$^{-1}$, most
of which occurs in emission lines.  
The flux in hot FUV emission lines from CTTSs increases with the mass
accretion rate \citep{Joh00,Cal04}.  Many \ion{Fe}{2} and fluorescent H$_2$ lines
are pumped by other emission lines, such as \Lya, so their strength should
also scale with the accretion rate.  \citet{Joh00} found that H$_2$ flux
scales with the \ion{C}{4} flux.  Other lines, such as the \ion{O}{1} 1305
\AA\ triplet and
the \ion{C}{2} 1335 \AA\ doublet, may be produced both in accreting gas and in the wind.  Thus,
we can analyze certain emission lines to probe the accretion flow.

For emission lines that are not corrupted by
wind absorption, we fit a single or multiple
Gaussians to the emission line profile to measure the line center, FWHM, and flux
in the line.  Tables 5 and 6 list parameters for identified atomic or
unidentified emission lines detected in the \STIS\ and
\FUSE\ spectra, respectively. For lines that have large wind absorption
components, we list only the continuum-subtracted flux detected in the
line.  Fluxes are not directly comparable between the \STIS\ spectrum and the \FUSE\ spectrum owing to temporal variability in the emission
(see Fig. 5).

Strong resonance lines, such as the \ion{C}{4} doublet at 1549 \AA, are
commonly detected in emission from late-type stars and are also strong in the
FUV spectrum of RU Lupi.
 We identify weak
emission lines using the atomic spectral line database of
\citet{Kur95}.   The identifications marked
by a '?' in Tables 5--6 are uncertain.  Since many lines,  including fluorescent
H$_2$ emission, \ion{Fe}{2}, and \ion{Cr}{2}, are pumped by \Lya, our
identification technique relies on 
 searching for lines that could be pumped by \Lya\ or other strong
emission lines.  We list the pumped lines and the pumping transition in Table 7.  Many of these lines were identified by \citet{Joh84},
\citet{Car88}, and \citet{Har00}.
Other pumped lines include \ion{Cl}{1} 1351.6 \AA, which is pumped by
\ion{C}{2} 1336 \AA, and \ion{S}{1} 1472 \AA, which is pumped by
\ion{O}{1} 1302 \AA\ and \ion{S}{1} 1295 \AA.  The \ion{O}{1} triplet at 1304 \AA\ can be pumped by Ly$\beta$ 1026 \AA\ \citep{Car88}.
Although airglow Ly$\beta$ prevents us from detecting the center of
Ly$\beta$ in our \FUSE\ spectrum, we do not detect Ly$\beta$ emission at
$100$ \kms.  Nevertheless, the \ion{O}{1} lines are very strong.  Moreover,
other lines such as \ion{C}{2} that require similar excitation are also
very strong.  We speculate that the \ion{O}{1} emission in the RU Lupi
spectrum is most likely produced by collisional excitation or recombination
in the hot gas rather than by fluorescence.

Many narrow (FWHM$\approx17.5$ \kms) emission lines detected in the \STIS\ spectrum are identified as
H$_2$ lines (see Table 8) based on the H$_2$ transition database
calculated by \citet{Abg93} and the linelist from \HST/STIS observations of
TW Hya \citep{Her02}.  Most of these lines are Lyman-band (B-X) H$_2$ lines pumped by \Lya.  We also detect one Werner-band (C-X) H$_2$ line at 1208.8 \AA\ that is pumped by \ion{O}{6} emission.
These H$_2$ lines are found in UV spectra of CTTSs
\citep{Bro81,Val00,Ard02a}, red giants \citep*{Ayr03a}, and the wind of Mira
B \citep{Woo02}.  We coadded the strongest H$_2$ lines in the STIS spectrum
to analyze the line profiles.  Figure 12 shows that the coadded H$_2$ lines
are well fit by a strong Gaussian, centered at -12 \kms, and a weaker
Gaussian, centered at -30 \kms, relative to the radial velocity of RU Lupi.
We also detect additional flux in these lines from $v=-120$ to $-70$ \kms.
The flux ratio of these two Gaussians is 1.3 and the ratio of the widths is
0.33.  To measure fluxes, we fit each individual H$_2$ line with two
Gaussians, with the velocity separation, relative width, and relative flux
fixed to the numbers obtained from the fit to the coadded profile above.
We also detect unresolved
Lyman and Werner-band H$_2$ lines with \FUSE.  
We fit a single Gaussian profile to these
lines, and list their fluxes in Table 9.  
If present, any extended H$_2$ emission would be detected by {\it FUSE}
because it has a  large ($30\farcs\times30\farcs$) aperture.
Figure 13 compares the coadded spectral line profiles of the H$_2$ lines in
the {\it STIS} and {\it FUSE} spectra, and the geocoronal Ly$\beta$ line in
the {\it FUSE} spectrum.  The {\it FUSE} H$_2$ lines are not significantly
broader than the {\it STIS} H$_2$ lines, convolved to the lower resolution
of {\it FUSE}.  If the H$_2$ emission were extended beyond a point source
in the dispersion direction, we would expect the spectral line profile
would be much wider.  This comparison therefore suggests that the H$_2$
emission in the {\it FUSE} spectrum is not significantly extended beyond a
point source.
The H$_2$ lines will be discussed 
in detail in a future paper.

Some cooler emission lines typically found in stellar chromospheres, such as the
\ion{O}{1} 1304 \AA\ triplet and the \ion{C}{2} 1335 \AA\ doublet, are
partially or completely absorbed by wind features listed in Table 10 (see
also Fig. 16).   Figure 14 shows that emission in these cool lines are centered at the radial velocity of RU Lupi.

Emission in \ion{C}{4}, \ion{Si}{4}, \ion{N}{5}, and possibly \ion{O}{6} is most likely produced by
$10^5$ K gas heated by the accretion shock.  Figure 15 shows that these lines are strong and broad.  The STIS lines have centroids 
near the radial velocity of the star.  The \FUSE\ lines of \ion{O}{6} 1032 \AA\ and \ion{C}{3} 977 \AA\
lines are blueshifted by $\sim 190$ and $\sim 100$ \kms, respectively.  Low S/N prevents any detection of wind in the \ion{C}{3} 977 \AA\ line, but
if present we may be underestimating the blueshift.  Absorption in H$_2$
and other lines may mask some redshifted emission in the \ion{O}{6} and
\ion{C}{3} lines.  The velocity shift of these lines may be
time dependent.  Alternatively, the emission in \ion{C}{3} and \ion{O}{6}
resonance lines during the \FUSE\ observation might be produced by shocked
emission due to the outflow, as hot gas can be seen in some HH objects
\citep{Ray97}.  The lack of \ion{C}{3} 1175 \AA\ emission suggests that any
emission produced by accreting gas is faint during this observation. \FUSE\
would then only detect this spatially extended blueshifted emission.  The \STIS\
observations would most likely not be sensitive to this extended emission
because of its small aperture.
The time dependence in the strength of
\ion{C}{3} emission is discussed in \S 7.

The \ion{C}{4} 1548 \AA\ line
shows broad blueshifted emission, extending to -300 \kms, that is not
present in the other emission lines.  
In contrast, the \ion{Si}{4} 1394 \AA\ line has stronger redshifted
emission than the \ion{Si}{4} 1403 \AA\ line.  The difference in behavior
between the \ion{C}{4} doublet and the \ion{Si}{4} doublet is puzzling.  
The \ion{N}{5} 1243 \AA\ line is absorbed by \ion{N}{1} in the wind (Fig
16), while interstellar H$_2$ absorption corrupts the \ion{O}{6} line
profiles (Fig. 8).  Anomalously weak \ion{N}{5} 1243 \AA\ emission was also detected in the FUV spectrum of T Tau \citep{Wal03}.  

Optical depth effects likely  produce these complex emission features.  The
mass accretion rate is large, so we infer that the accretion column may be
optically thick.  The \ion{C}{4} 1548 \AA\ and 1550 \AA\ lines from TW Hya
have a similar shape to each other \citep{Her02}, possibly because the mass
accretion rate of TW Hya is 25 times smaller than that of RU Lupi, so the
accretion shock may be less optically thick.
We also note that, unlike TW Hya, \ion{Si}{2}, \ion{Si}{3}, and \ion{Si}{4} lines from RU Lupi are all bright.  Understanding these profiles will require detailed modelling of the accreting gas.

Most of the observed \Lya\ emission is absorbed by \ion{H}{1} in
the stellar outflows, in the ISM, and possibly the accretion shock itself.  In \citet{Her04}, we found that the intrinsic \Lya\ emission could be reconstructed from the H$_2$ lines, and may contribute 85\% of the total FUV emission from CTTSs.  The strength of the H$_2$ and \ion{Fe}{2} lines that are pumped by \Lya\ imply that the \Lya\ emission from RU Lupi is similarly strong.  We will estimate the \Lya\ emission from RU Lupi in a future paper.

Coronal Fe lines have been detected in UV spectra of the Sun \citep*{San77}
and cool stars \citep{Ayr03,Red03}.  \citet{Lam01} predict that emission in the
[\ion{Fe}{12}] 1349.382 \AA\ and \ion{Fe}{11} 1467.420 \AA\ 
lines may be produced at the accretion shock and could be 
detected in UV spectra of CTTSs.  Based on models of the 
accretion shock by \citet{Lam01}, the lines should be centered 
at $v_{shock}/4$.  The FWHM of these emission lines should be dominated by 
thermal broadening.    We do not detect any coronal Fe lines in the RU
Lupi spectra, and place unreddened upper limits of $\sim2\times10^{-15}$ \erg\ that are about
an order of magnitude larger than the line fluxes predicted by
\citet{Lam01}. 

\section{WIND ABSORPTION}
Outflows from RU Lupi are detected by the strong blueshifted absorption seen in
P-Cygni profiles and against the continuum (Fig. 6).  The absorption is typically seen from low energy levels of neutral or singly-ionized species.

The cool emission lines that are not
corrupted by wind absorption are symmetric about
the radial velocity of the star (see \S 5.2 and Figure 14).  We therefore estimate the intrinsic
profile of lines with wind absorption by reflecting the red emission about the radial velocity of the star.  We then measure the equivalent width and central and maximum
velocities of the absorption feature (see Table 10).
Lines such as \ion{O}{1}, \ion{C}{2}, and \ion{Fe}{2} have P-Cygni profiles, with
absorption ranging from 
about $-70$ to $-270$ \kms.  For certain lines of \ion{Fe}{2} and
\ion{N}{1}, the wind absorption is detected against the continuum.
The wind absorption feature seen in \ion{Si}{3} is narrower than the wind
absorption features detected in other lines from the ground state, which may be due either to a small
column density of \ion{Si}{3} or the absence of a velocity gradient for the 
hot gas in our line
of sight.  The latter possibility could occur if most of the \ion{Si}{3} occurs close to the star.  
Most of the wind absorption is detected in transitions with low-energy lower levels.
However, we also detect wind absorption in the 
\ion{N}{1} lines at 1242, 1492, and 1494 \AA\ from lower levels excited to 2 eV
(Fig. 16).

In certain cases, the wind absorption is strong enough that many nearby
lines are barely detected.  Figure 17 shows the \ion{O}{1} triplet at 1304
\AA\ and the \ion{C}{2} doublet at 1335 \AA.  The wind in the \ion{C}{2}
1335.7 \AA\ line obscures emission in \ion{C}{2} 1334.5 \AA.  Similarly, the \ion{O}{1} 1305 \AA\ line is weak due to wind in that line and a wind in \ion{O}{1} 1306 \AA.

\section{VARIABILITY}
Figure 5 shows a striking decrease in flux in the \ion{C}{3} 1175 \AA\
multiplet between the \STIS\ and the \FUSE\ observations.  In the \FUSE\
spectrum, most of the emission detected near 1175 \AA\ is H$_2$ and not
\ion{C}{3}.  The \ion{O}{6} 1035 \AA\ doublet and \ion{C}{3} 977 \AA\
emission are weak and blueshifted.  If the drop in flux seen in \ion{C}{3}
1175 \AA\ is related to a decrease in the mass accretion rate, then we may
also not be detecting any emission produced by the accretion shock in the
\ion{}{6} 1035 \AA\ doublet or the \ion{C}{3} 977 \AA\ line.  Instead, with 
its large aperture, \FUSE\ may have detected emission in these lines that
is produced by shocks in the outflow.
The fluorescent H$_2$ emission in the \FUSE\
spectrum, which is highly sensitive to the \Lya\ emission, is comparable to
the flux we expect based on the \STIS\ spectrum.  In \S 5.2, we found that
the H$_2$ emission in the \FUSE\ bandpass originates predominantly
on-source.  Because the H$_2$ lines serve as a proxy for the strength of
\Lya\ emission, we infer that the \Lya\ emission has a similar flux in the
two observations.  This result is surprising because we expect that both
\ion{C}{3} and \Lya\ emission are produced by accreting gas.  We speculate
that a smaller mass loss rate significantly reduces the optical depth of
the \ion{C}{3} line, while the \Lya\ line remains optically thick.
The FUV continuum where the STIS and \FUSE\ spectra overlap
is marginally detected at $\sim 2\times10^{-15}$ \erga.  This emission, which is likely produced by H$_2$ \citep{Ber04}, did not change by more than $\sim 50\%$.

We also compare the observed STIS spectrum with spectra obtained using
the G160M grating on
\HST/GHRS, which cover narrow (35 \AA) spectral regions centered 
on the \ion{C}{4} and \ion{Si}{4} lines \citep{Ard02a}.  In the GHRS
spectra, the FUV emission lines have similar shapes but are four times weaker than the fluxes
detected with \STIS\ (Fig. 18).  The continuum in the GHRS
observation also appears weaker than that detected in the STIS observation.
However, the H$_2$ emission line fluxes
are similar in the \STIS\ and the GHRS spectra, which suggests 
that the \Lya\ emission did not change significantly between the two
observations.  The GHRS observation could also include off-source H$_2$
emission, as is detected towards T Tau \citep{Wal03}.  Our \STIS\ observations must have fortuitously occurred when the accretion rate of RU Lupi was higher than usual.

We find no significant variability during the \STIS\ spectrum.  Although this analysis is complicated because the count rate depends on the changing focus of {\it HST} (see \S 3.1), we place a limit that the flux of RU Lupi varied by no more than 10\% in intervals longer than 100 s and 4\% in intervals longer than 300 s.
Some minor differences appear in the profiles of hot lines during the
\STIS\ observations.
The S/N of the \FUSE\ observations prevented any significant
analysis of variability over short timescales.

\section{CONCLUSIONS}
We present and analyze FUV spectra of RU Lupi, a 2--3 Myr old CTTS with a
high mass accretion rate.  Based on these spectra, we
find the following results:

1. The \HST\ telescope focus improved during each orbit of the RU Lupi observation, which increased the detected count rate because the FWHM of the line-spread function was larger than the aperture size in the dispersion direction.  We account for this instrumental problem by using observations of V471 Tau.  This problem appears in several other time-tag STIS spectra that used a small aperture.

2. We estimate a total hydrogen column density $\log N$(H)$=20.1\pm0.2$ in our line of sight to RU Lupi using Ly$\alpha$ and H$_2$ absorption, and convert the column density to an extinction of $A_V\sim0.07$ mag.

3.  The mass accretion rate of RU Lupi is about $(5\pm2) \times 10^{-8}$
$M_\odot$ yr$^{-1}$.

4.  Emission is detected in 174 lines, 103 of which are Lyman and Werner-band
H$_2$ lines pumped by \Lya\ or \ion{O}{6}.  Some \ion{Fe}{2} and \ion{Cr}{2} lines are also
pumped by \Lya.   The hot lines of \ion{C}{4}, \ion{Si}{4}, and \ion{N}{5}
are bright and broad, but differences in the line profiles are puzzling.
In particular, the \ion{C}{4} 1548 \AA\ line has strong redshifted emission 
that is not seen in the \ion{C}{4} 1550 \AA\ line, but the \ion{Si}{4} 1394 \AA\
line has strong blueshifted emission relative to the \ion{Si}{4} 1403 \AA\
line.  Emission in the \ion{O}{6} 1035 \AA\ doubet and \ion{C}{3} 977 \AA\
is blueshifted, and at the time of our \FUSE\ observation may be dominated
by emission produced by shocks in the outflow.  The observed \Lya\ emission
only accounts for about 1\% of the observed FUV emission, likely because of
strong absorption by \ion{H}{1} in the ISM and stellar wind.

5. Wind absorption is detected in 30 different lines, and generally ranges from
-70 to -270 \kms.  Atoms with low first-ionization potentials are ionized in
the wind, resulting in absorption in
lines of \ion{Si}{2}, \ion{Si}{3}, \ion{Fe}{2}, \ion{Al}{2}, and
\ion{C}{2}.  Wind absorption is also detected in lines of \ion{N}{1} and
\ion{O}{1}.  Three \ion{N}{1} wind absorption lines originate from levels excited 2.38 eV
above the ground state.

6.  The FUV emission from RU Lupi varies significantly over months-long
timescales.  Our \STIS\ observation of RU Lupi occurred  during
a period of strong emission, possibly due to a high mass accretion rate.  The \ion{C}{3} emission is much weaker in the \FUSE\ spectrum than in the \STIS\ spectrum, although the H$_2$ emission, which is a proxy for \Lya\ emission, does not vary significantly between these two observations.

\section{ACKNOWLEDGEMENTS}

This research was funded in part by STScI grant GO-08157.01-97A to SUNY
Stony Brook, by STScI grant GO-08157.02-97B to the University of Colorado,
NASA contract NNGO4GJ31G from Johns Hopkins University to the Unversity of
Colorado, and by the Swedish National Space Board. This paper is based on
observations made with the NASA/ESA Hubble Space Telescope, obtained at the
Space Telescope Science Institute, which is operated by the Association of
Universities for Research in Astronomy, Inc., under NASA contract
NAS5-26555, and on observations obtained by the NASA-CNES-CSA FUSE
mission.
FUSE is operated for NASA by the Johns Hopkins University under NASA contract NAS5-32985.
These observations are associated with {\it HST} program 8157 and {\it FUSE} program A109.  
We thank the anonymous referee for insightful comments that 
improved the paper.
We also thank Paul Goudfrooij for suggestions and discussion concerning the
instrumental problems with {\it HST}/STIS.  GH thanks Brian Wood and Seth
Redfield for discussion related to the calibration of the STIS spectra.

% Bibliography

\begin{figure}
\plotone{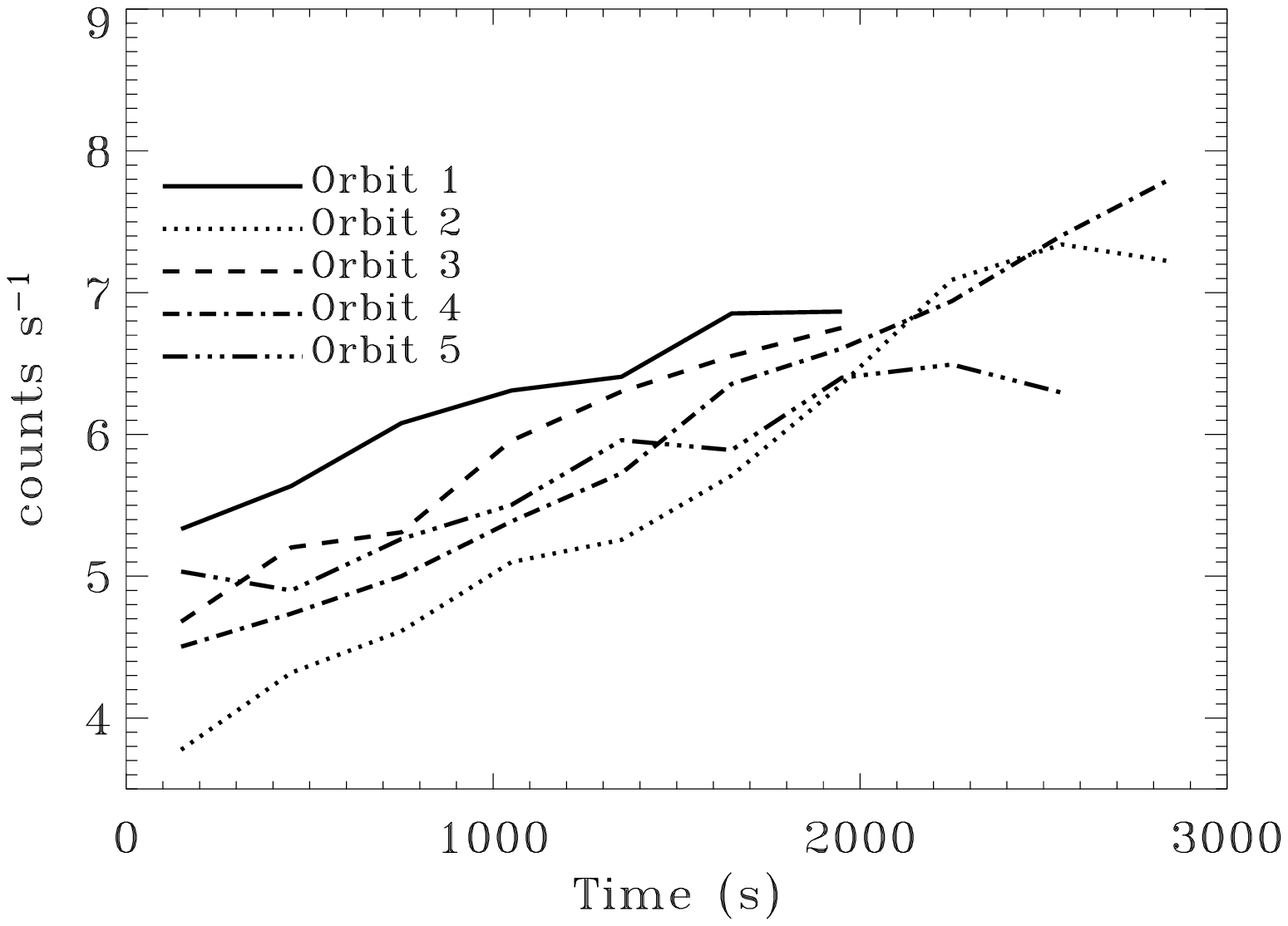}
\caption{In our STIS observations of RU Lupi, the count rate within each
orbit increased with time, but then returned to a low level at the start
of the next orbit.  This increase of flux results because the FWHM of
the line-spread function is wider than the aperture in the dispersion
direction but narrows during each orbit, due to thermal changes in 
the telescope.}
\end{figure}

\begin{figure}
\plotone{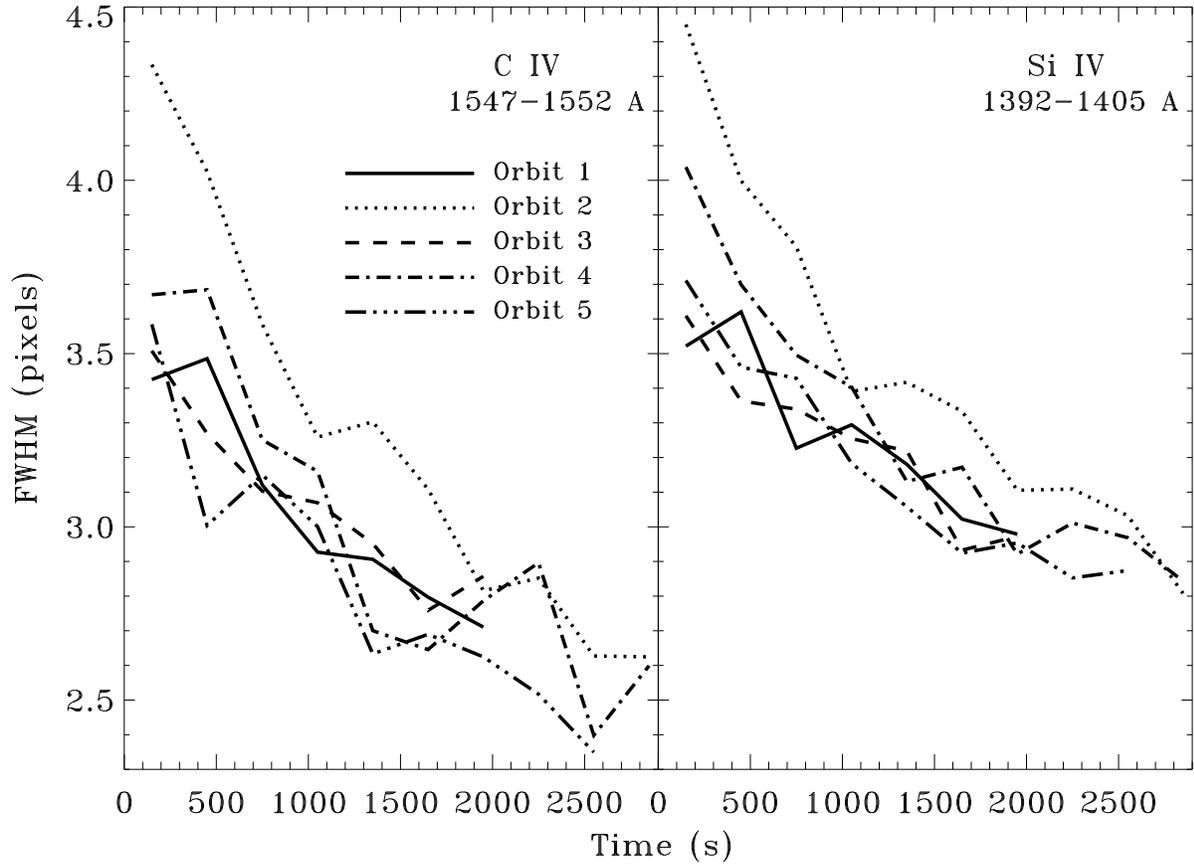}
\caption{The FWHM of \ion{C}{4} and \ion{Si}{4} emission in the
cross-dispersion direction, measured in 300 s intervals, improves during
each orbit because of thermal focus variations.}
\end{figure}

\begin{figure}
\plotone{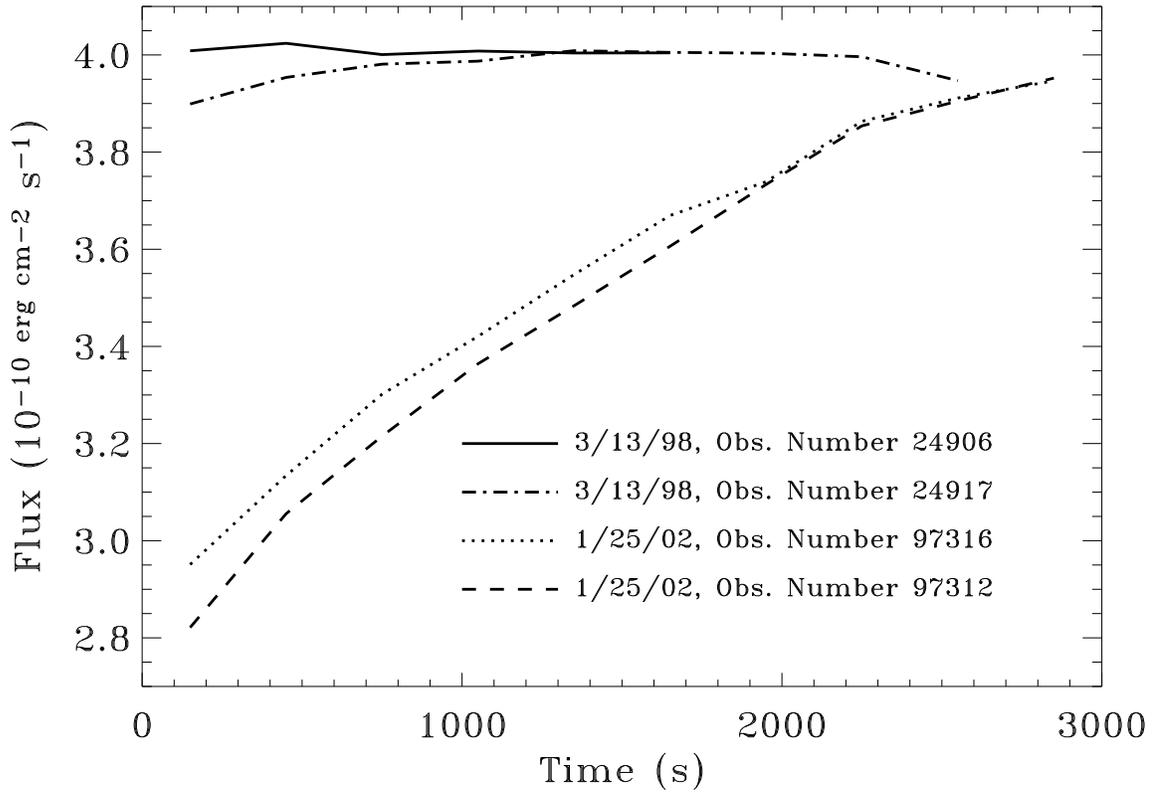}
\caption{The flux detected between 1250--1600 \AA\ in four of 16 orbits of STIS observations of V471 Tau.  A gradual increase of detected flux with time is present in two of the orbits shown here, and is similar to the increase seen in our spectra of RU Lupi.}
\end{figure}

\begin{figure}
\plotone{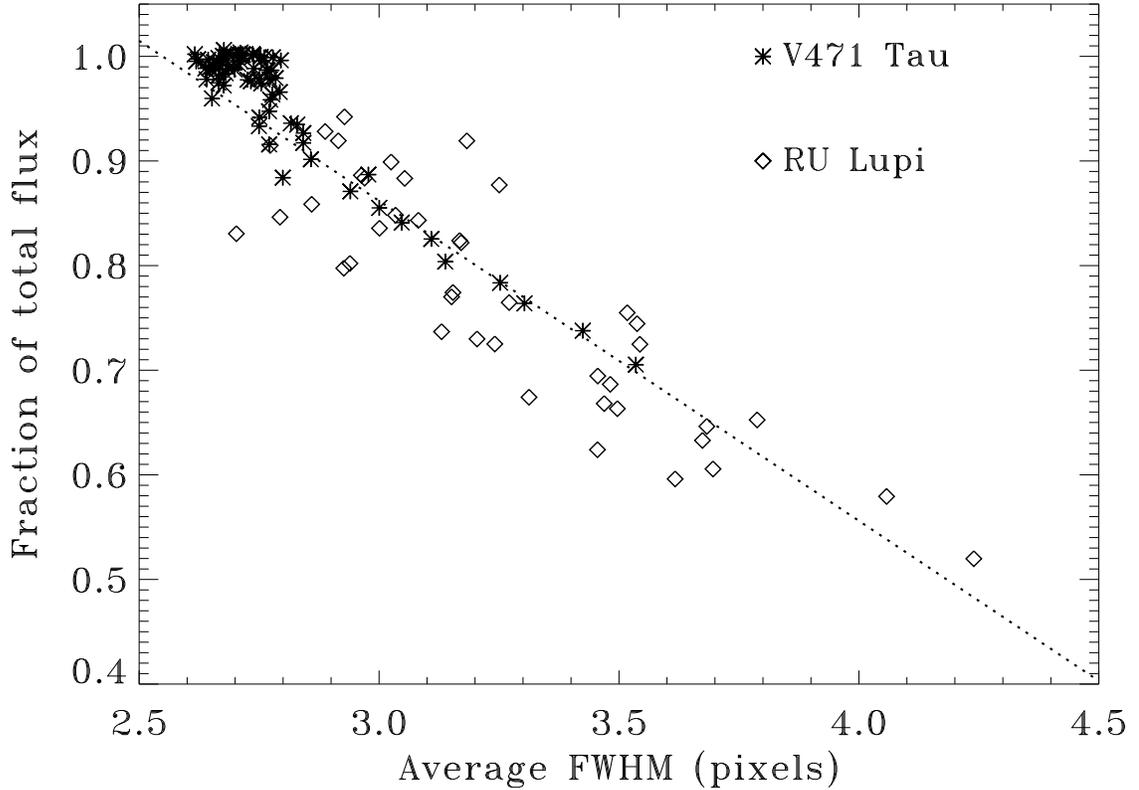}
\caption{The flux between 1230--1650 \AA\ from RU Lupi (diamonds) and V471
Tau (asterisks) depends on the telescope focus.  The fraction of the total
flux is calculated by comparing the measured flux in a 300 s interval to
the estimated total flux, for the wavelength region between 1230--1650 \AA.
The image size in the dispersion direction is plotted here as the average
of the FWHM of \ion{O}{1} (1292--1313 \AA), \ion{C}{2} (1333--1337 \AA),
\ion{Si}{4} (1392--1405 \AA), and \ion{C}{4} (1547--1552 \AA).  The dotted
line is the line that best fits the data from the calibration source V471
Tau.  We then estimate the total flux from RU Lupi by fitting the RU Lupi
data to this line using the least squares minimization method.
The point-spread function is broader than the small aperture, so that
better focus improves the count rate on the detector increases.}
\end{figure}

\begin{figure}
\plotone{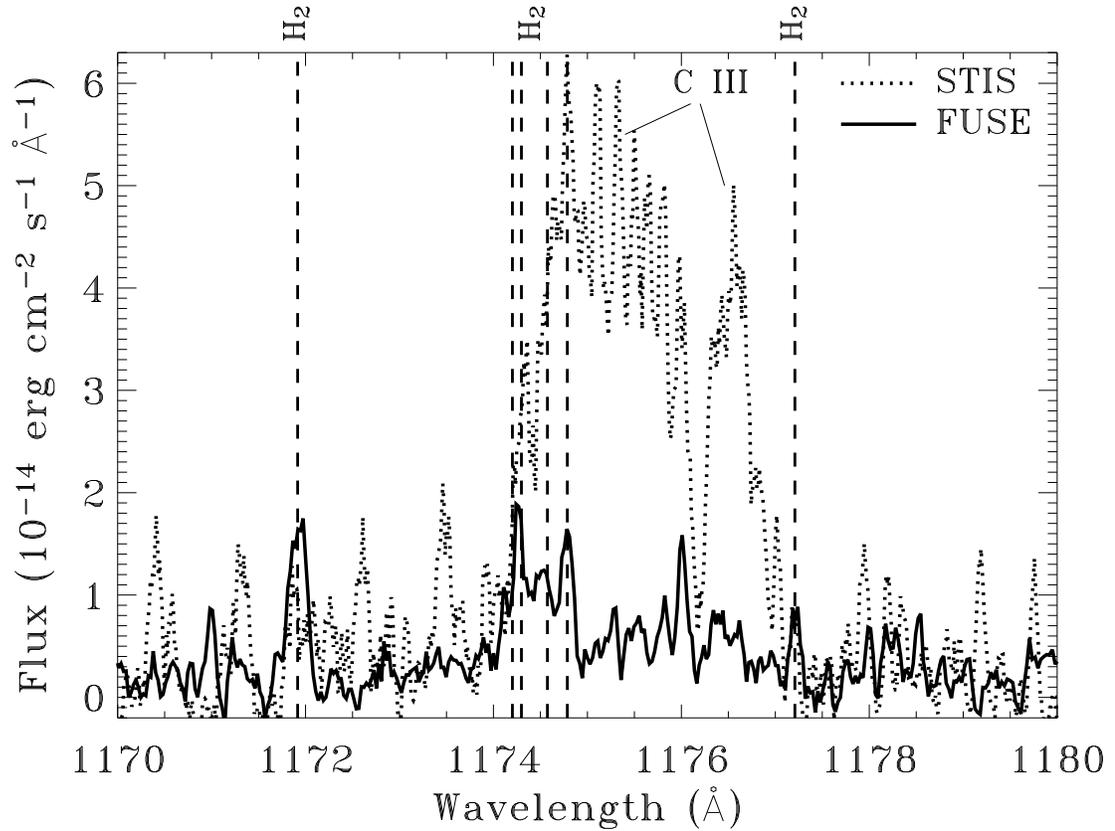}
\caption{\ion{C}{3} emission was strong during the HST/STIS observation
(dotted line), 
but was much weaker during the \FUSE\ observation (solid line), which occurred about a year
later.  This striking difference likely resulted because of a drop in the
mass accretion rate between the two observations.  The wavelengths of fluorescent H$_2$ emission lines are indicated by the dashed vertical lines.}
\end{figure}

\begin{figure}
\plotone{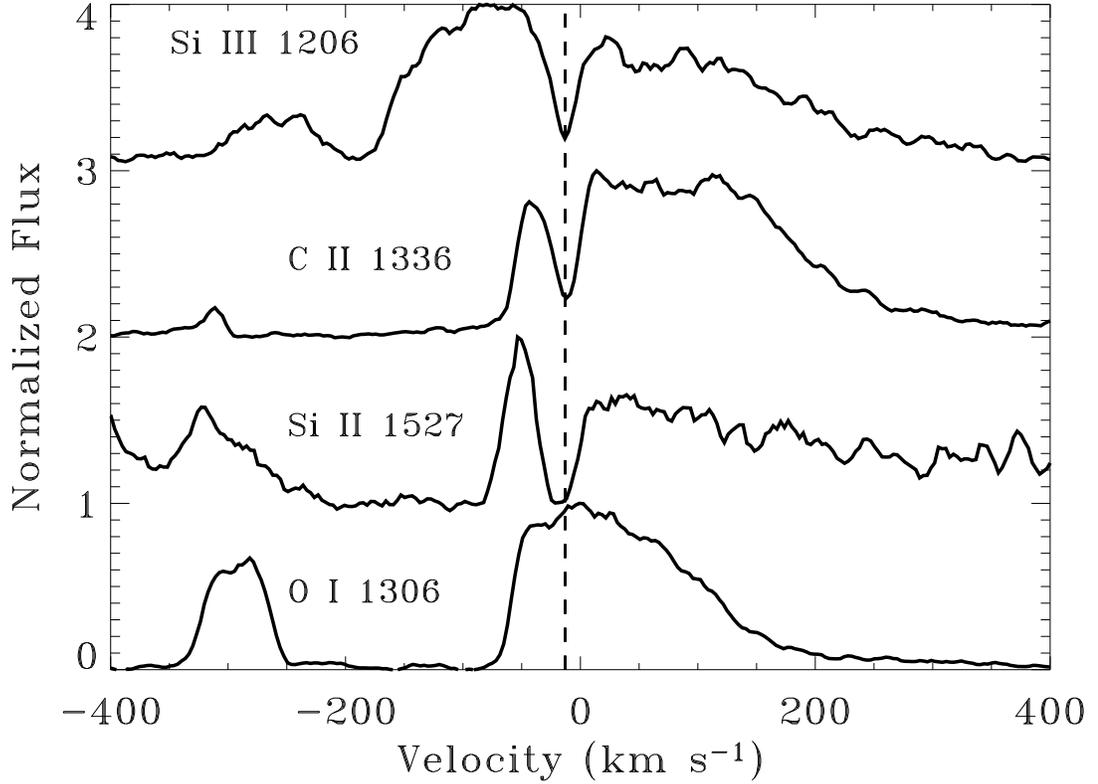}
\caption{P-Cygni profiles are detected for many lines, four of which are
shown here.  The wind absorptions in most lines range from $\sim-70-270$
\kms, although the absorption in the \ion{Si}{3} line is much narrower than wind absorption from ground state transitions of other species 
because the line is less optically thick.
Narrow absorption near line center (except for the excited \ion{O}{1} 1306 \AA\ line) is produced by the interstellar medium
and possibly also by material local to RU Lupi.  Interstellar \ion{Si}{3} absorption is
not detected locally \citep{Woo02b}.  The dashed vertical line marks the average velocity of interstellar absorption detected towards RU Lupi.}
\end{figure}

\begin{figure}
\plotone{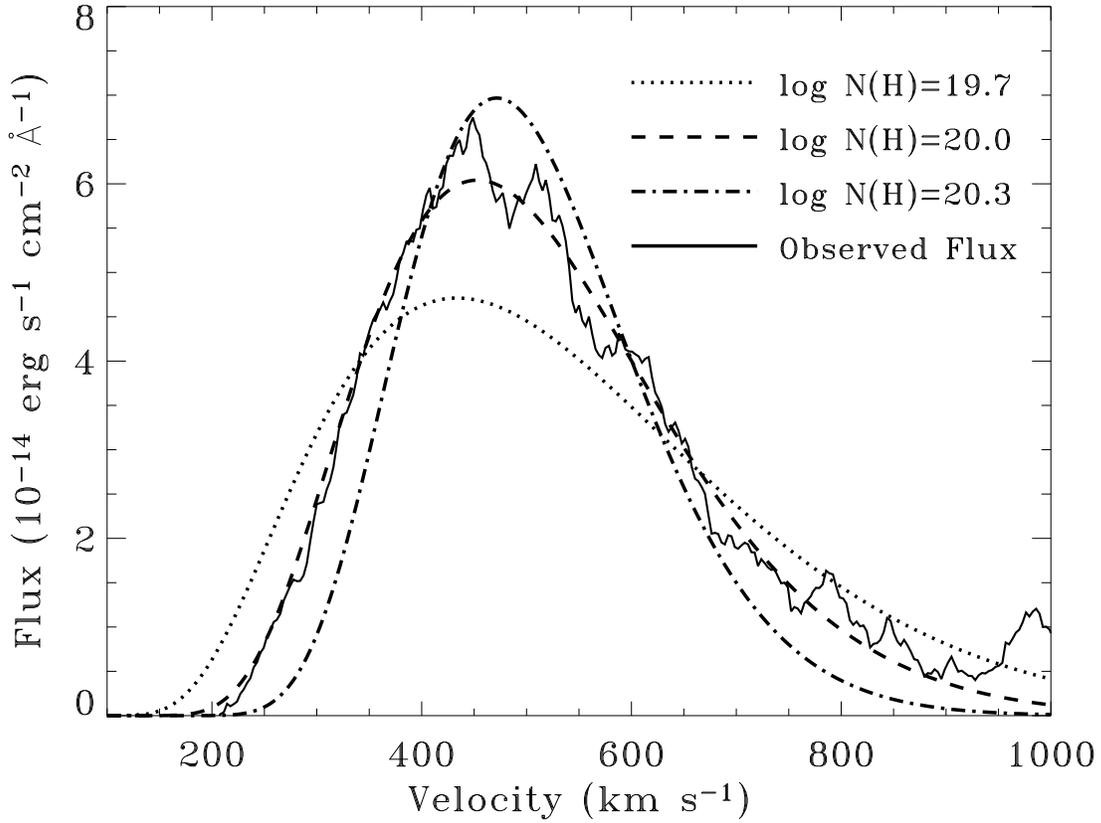}
\caption{The observed red wing of the \Lya\ line (solid line) is modeled as a Gaussian
profile, centered at the heliocentric velocity of \Lya, and a Voigt
absorption profile for interstellar \ion{H}{1}, centered at the
interstellar velocity measured from other absorption lines (see Table 5).  No \Lya\ emission is detected on the blue side of line center.  We
show here the fit for three different assumed column densities of \ion{H}{1}.
We determine that $\log N$(\ion{H}{1})$\approx 20.0\pm0.15$.}
\end{figure}

\begin{figure}
\plotone{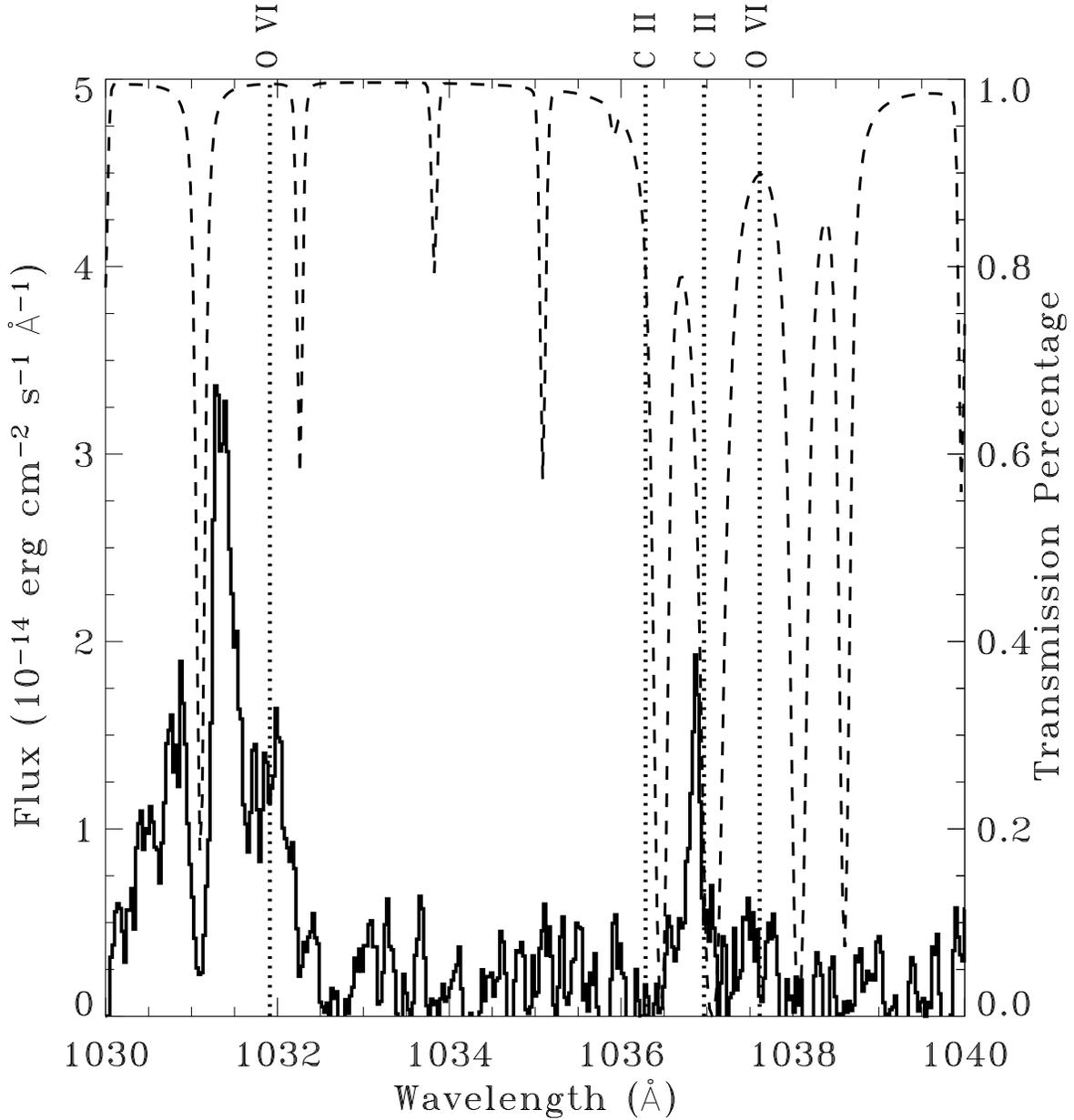}
\caption{The \FUSE\ spectrum of RU Lupi (solid line) is compared with the
transmission percentage (dashed line) through gas with $T=300$ K and
$\log N$(H$_2$)=19.0.  Both lines in the \ion{O}{6} doublet are blueshifted from line center (dotted
vertical lines).
The \ion{O}{6} 1032 \AA\ profile shows strong emission,
with absorption in the H$_2$  line at 1031.19 \AA.  The \ion{O}{6} 1037
\AA\ line is detected, but most of the line is absorbed by \ion{C}{2} (dashed
vertical lines) and H$_2$ in the line of sight to the star.}
\end{figure}

\begin{figure}
\plotone{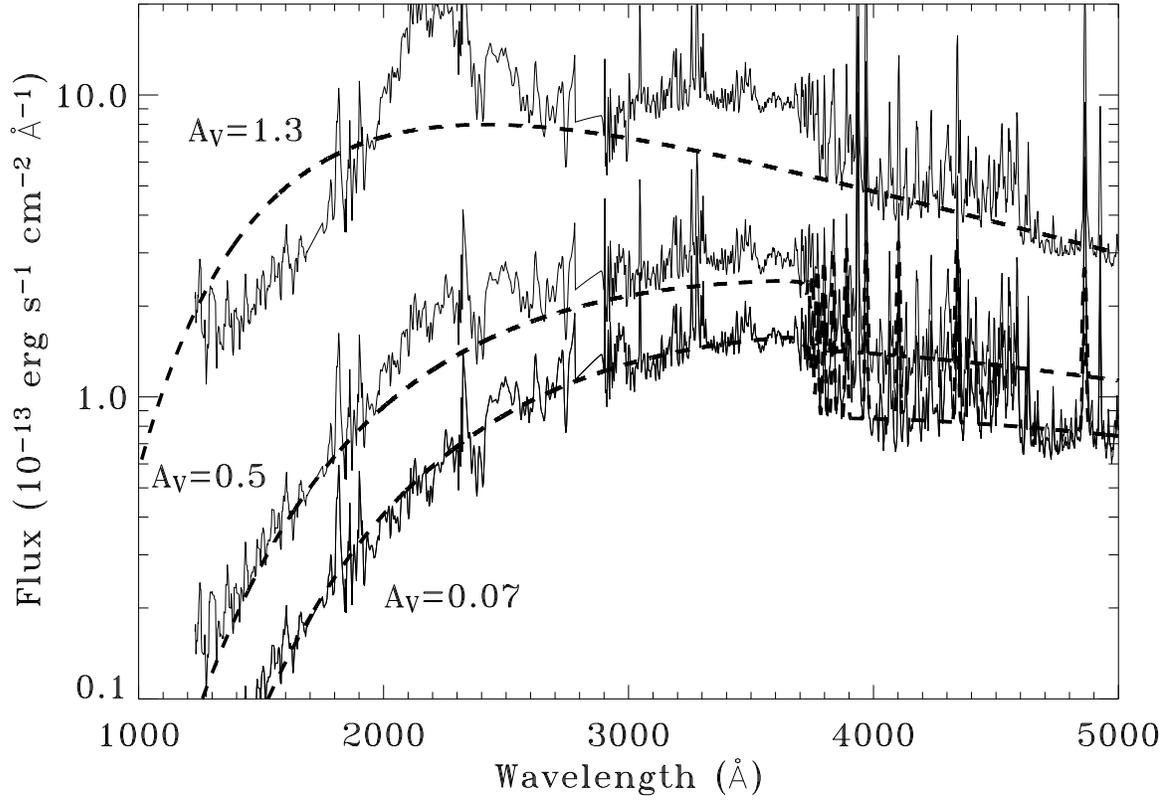}
\caption{We fit models (dashed lines) of the accretion continuum to the observed
UV-optical spectrum of RU Lupi, assuming a range of extinctions.  When we
assume $A_V>0.4$ mag, we cannot reconcile the increased emission at 2175 \AA\ with the emission at shorter wavelengths.  
Low extinction is also consistent with the observed $N$(\ion{H}{1}) column
density in our line of sight to the star (see \S 4.1).}
\end{figure}

\begin{figure}
\plotone{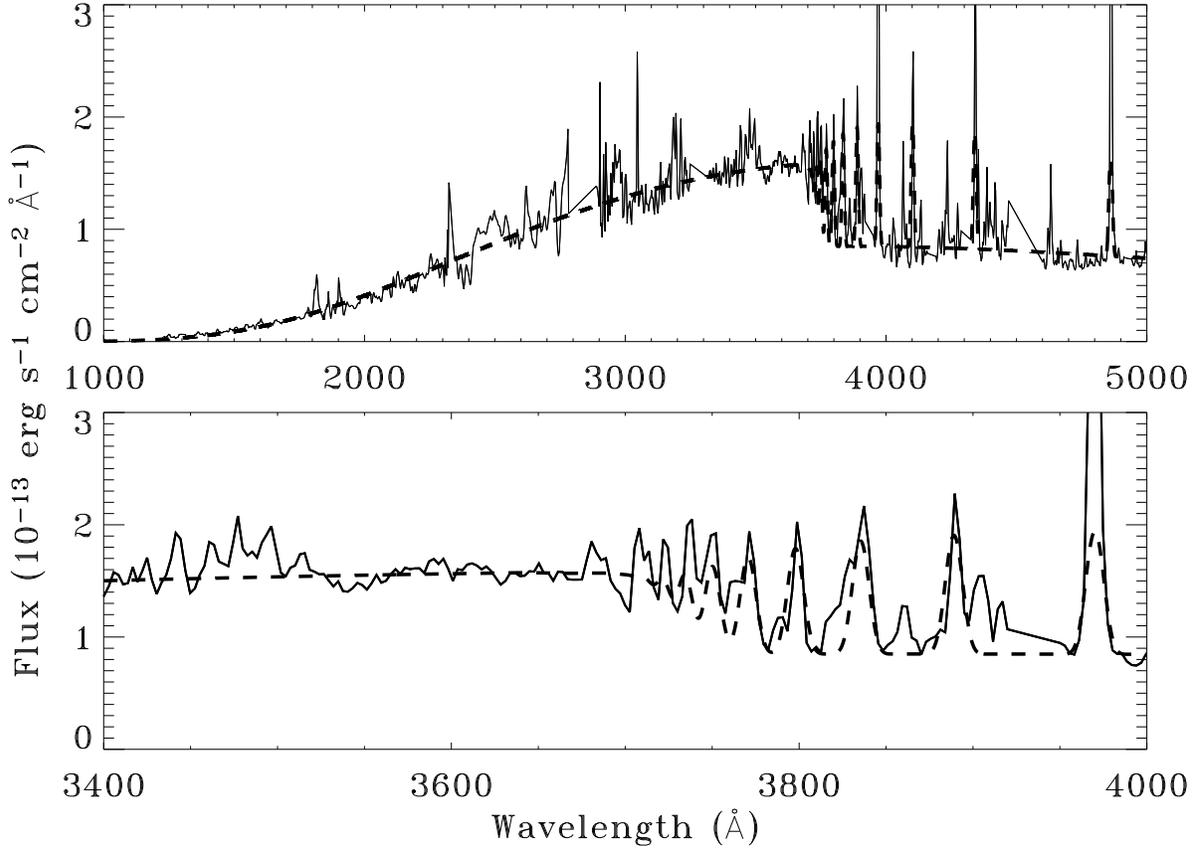}
\caption{We model the observed accretion continuum (solid line)  with
emission from a pure hydrogen slab (dashed line).  The accretion continuum
is obtained from data described in Table 3, assuming $A_V=0.07$.  To
extract the accretion continuum shown here, we subtracted the scaled
spectrum of the WTTS V819 Tau, a template for the photospheric emission of
RU Lupi, from the observed spectrum of RU Lupi.}
\end{figure}

\begin{figure}
\plotone{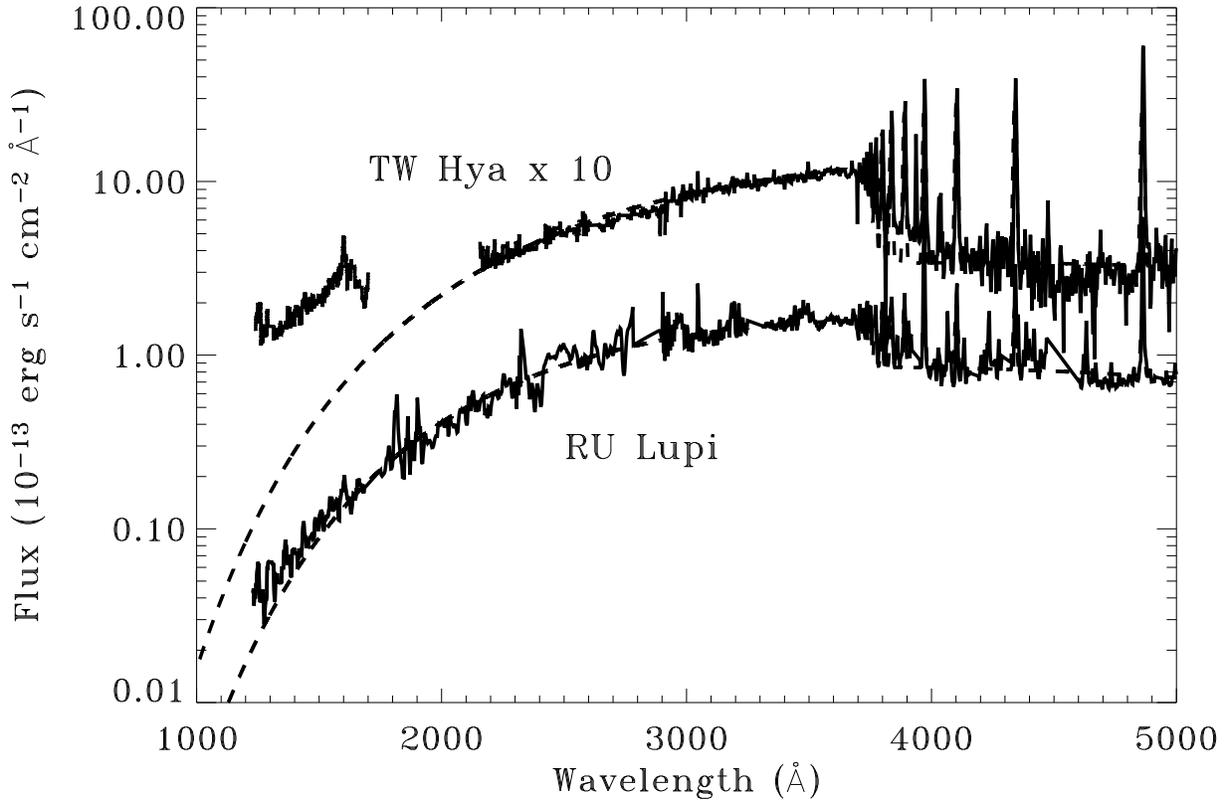}
\caption{Comparison of the observed UV emission from RU Lupi (bottom solid line) and TW Hya (top solid line, multiplied by 10) with spectra calculated from modelling the accretion continuum (dashed lines).  The FUV continuum in excess of emission from the accretion column is much weaker from RU Lupi than from TW Hya.}
\end{figure}

\begin{figure}
\plotone{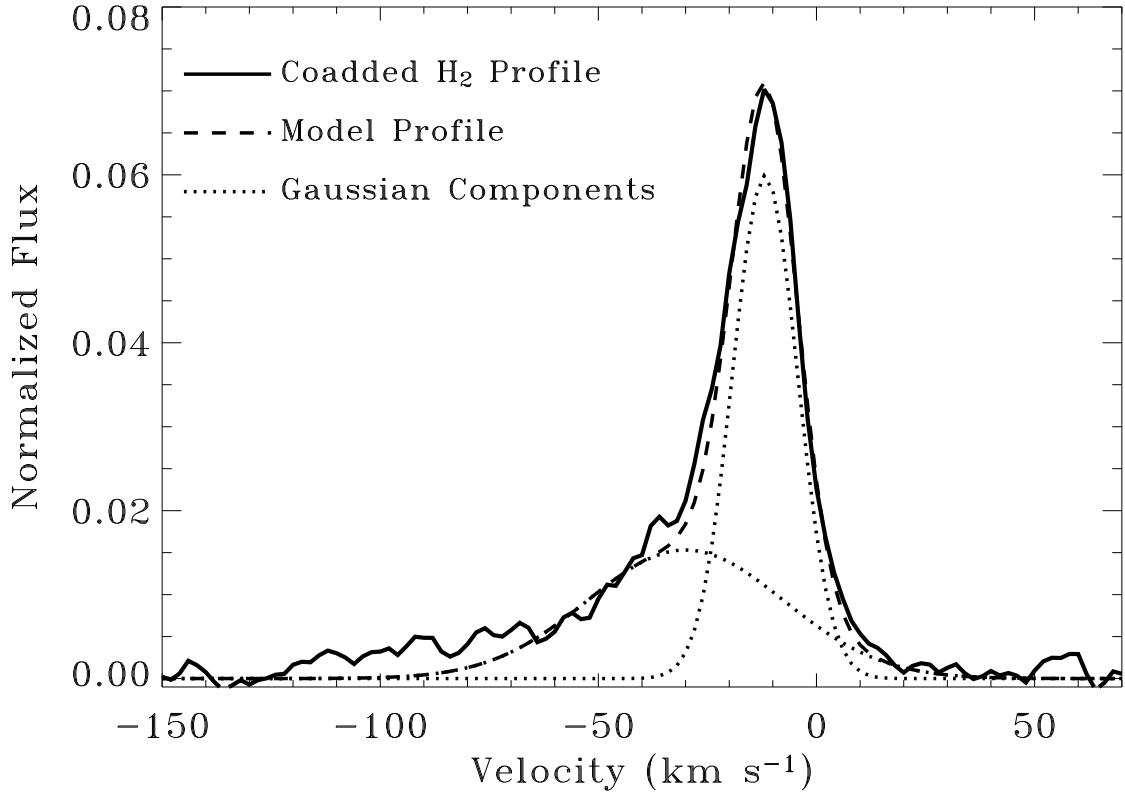}
\caption{The spectral profile of coadded H$_2$ lines from the STIS spectrum
(solid line) shows that the lines are blueshifted and asymmetric.  We fit
the profile with two Gaussian profiles (dotted).  The summed profiles are
shown by the dashed line, with parameters described in \S 5.2.  Excess
blueshifted emission at $v=-120$ to $-70$ \kms\ is also detected.  These
lines will be discussed in detail in a future paper.}
\end{figure}

\begin{figure}
\plotone{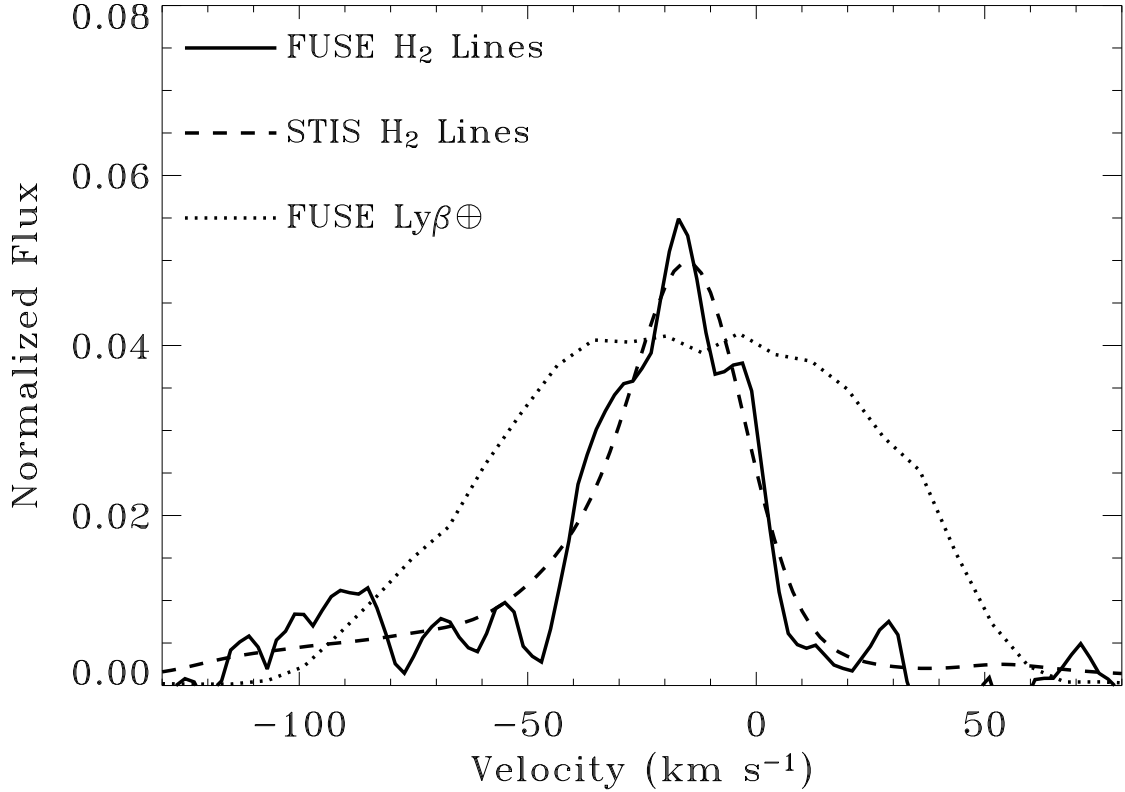}
\caption{We compare the spectral profile of coadded H$_2$ lines observed
with {\it FUSE} (solid line) with geocoronal Ly$\beta$ emission (dotted line)
and coadded H$_2$ lines from the STIS spectrum (dashed line), convolved to
the spectral resolution of {\it FUSE}.  The geocoronal Ly$\beta$ emission
fills the aperture and in this plot is shifted in velocity to the
center of the H$_2$ lines.  The coadded {\it FUSE} H$_2$ lines have a
similar FWHM as the coadded STIS H$_2$ lines, which suggests that the H$_2$
emission detected with {\it FUSE} is dominated by on-source emission.  Any
emission extended in the dispersion direction would broaden the line
profile.}
\end{figure}

\begin{figure}
\plotone{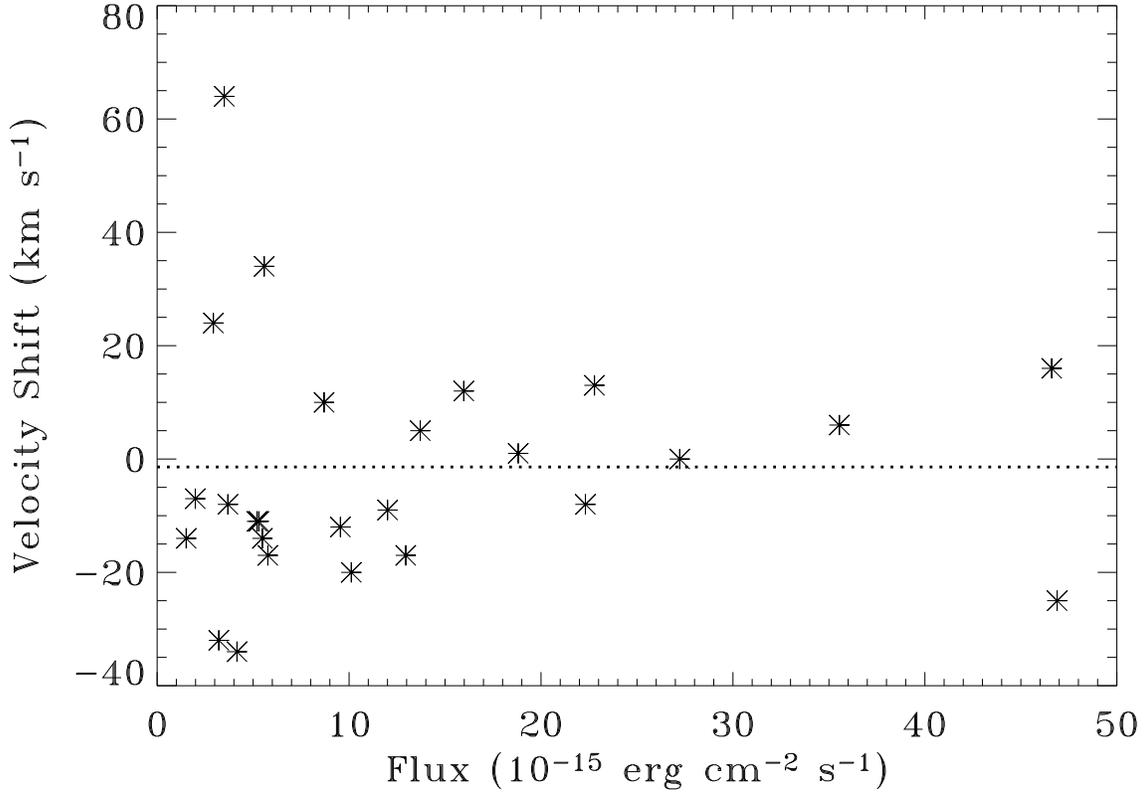}
\caption{The weighted mean of the radial velocity of cool (neutral, singly, and doubly ionized) lines is -1.4 \kms\ (dotted line) relative to the radial velocity of RU Lupi.  Lines that are contaminated by wind emission and lines that are only tentatively identified (marked by a ? in Table 5) are not used in this analysis.}
\end{figure}

\begin{figure}
\resizebox{4.5in}{6.5in}{
\plotone{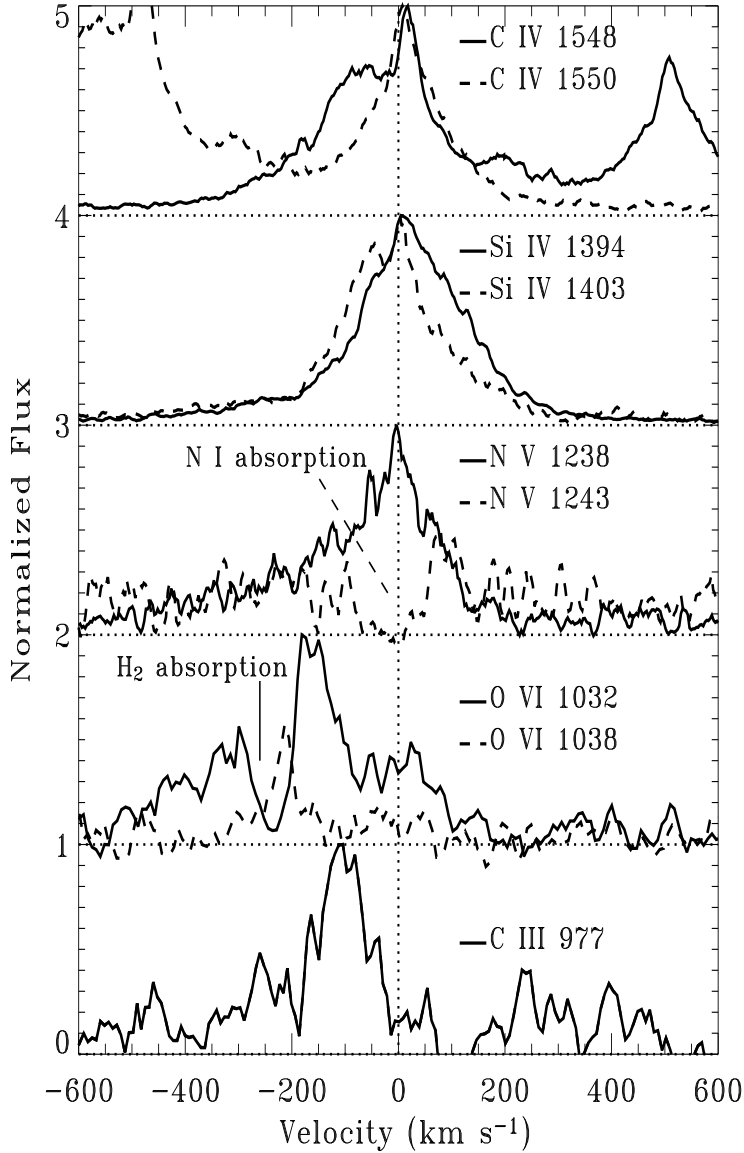}}
\caption{Emission lines produced in $\sim10^5$ K gas are broad and strong
in spectra of CTTSs.  These lines from RU Lupi show complicated features.
The \ion{C}{4} 1548 \AA\ line shows a strong blue bump that is not detected
in any of the other lines, while the \ion{Si}{4} 1394 \AA\ line shows
excess redshifted emission relative to the other lines.  \ion{N}{5} 1243
\AA\ emission is absorbed by \ion{N}{1} in the wind (see Fig. 16).  The
\ion{O}{6} and \ion{C}{3} lines detected in the {\it FUSE} spectrum appear
blueshifted and may be produced by shocks in the outflow.  Because the \FUSE\ observation occurred about one year after the \STIS\ observation and the FUV emission from RU Lupi is variable, the \ion{O}{6} and \ion{C}{3} lines shown above should not be directly compared with the \ion{C}{4}, \ion{Si}{4}, or \ion{N}{5} lines.}
\end{figure}

\begin{figure}
\plotone{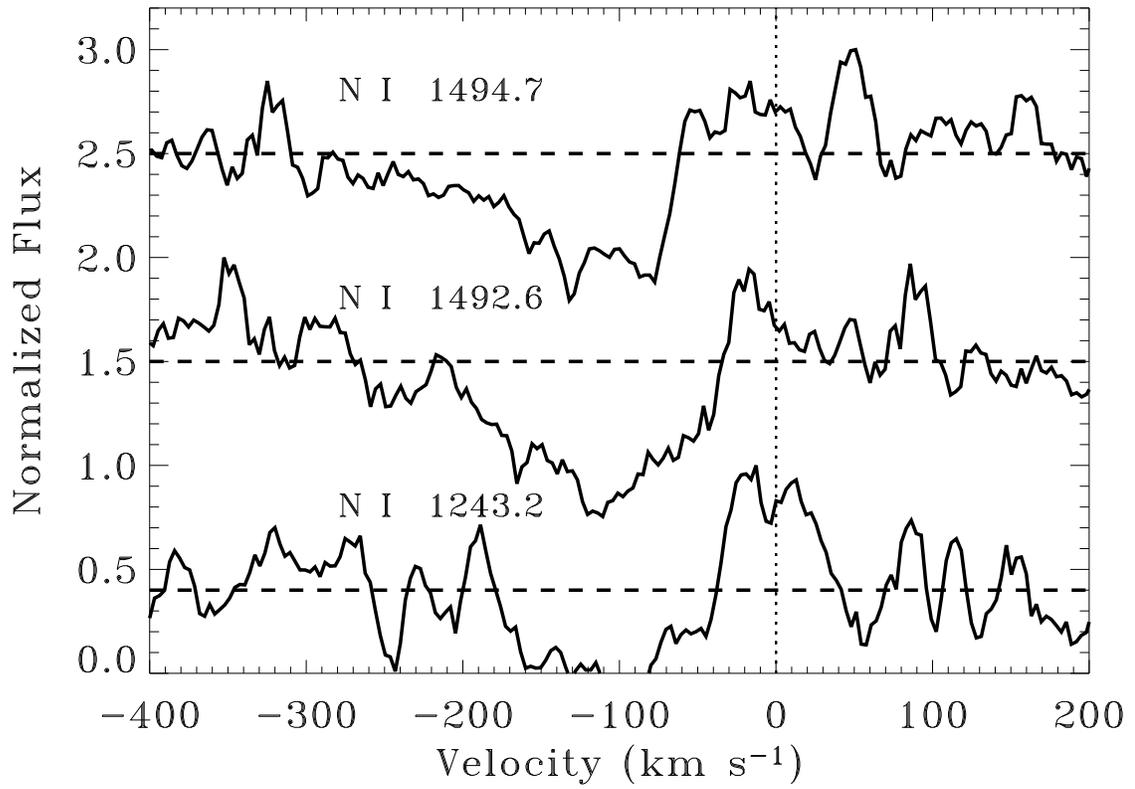}
\caption{Wind absorption is detected in \ion{N}{1} lines from excited lower
levels, with E=2.38 eV.  The \ion{N}{1} absorption near 1243 \AA\
attenuates the \ion{N}{5} 1243 \AA\ emission line.}
\end{figure}

\begin{figure}
\plotone{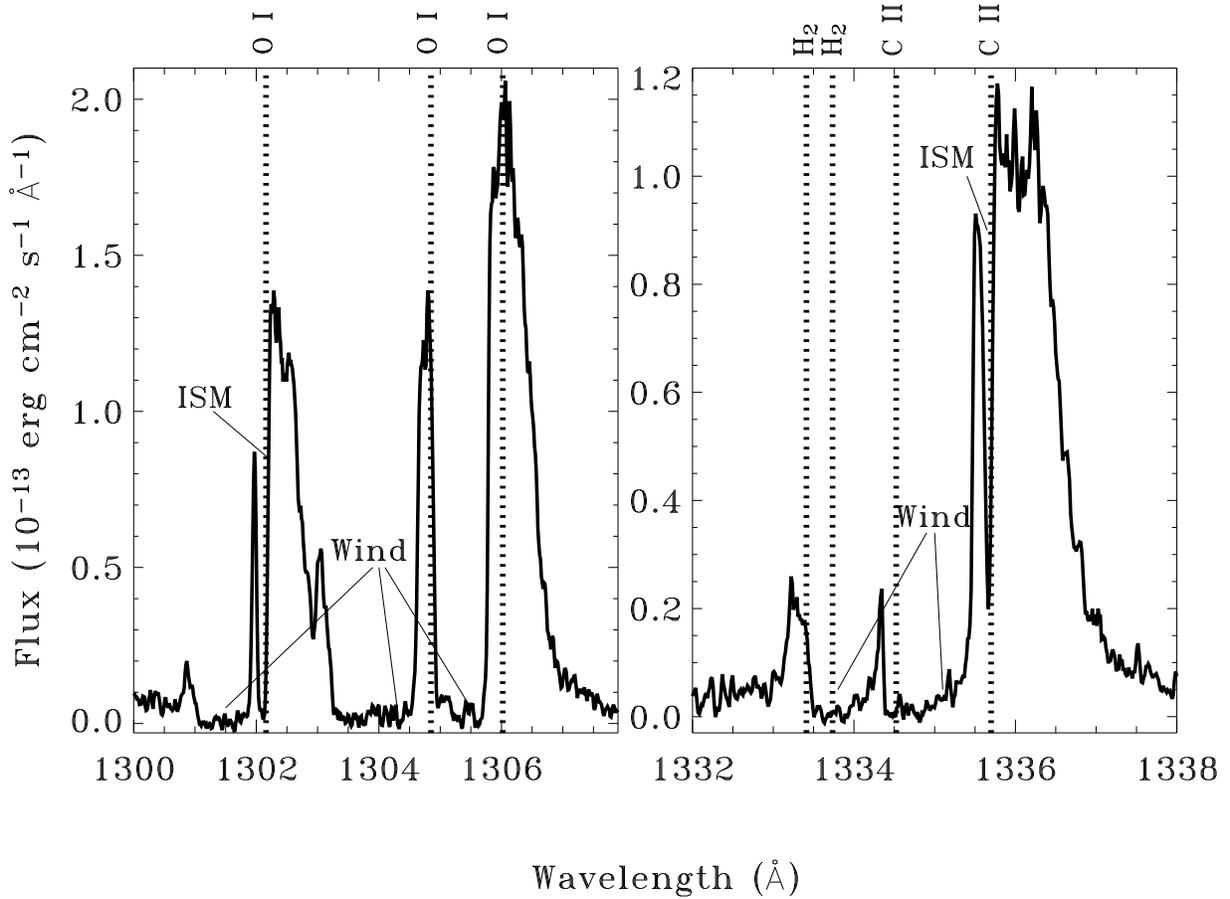}
\caption{Wind absorption in the \ion{O}{1} and \ion{C}{2} lines.  The wind absorbs a significant amount of the emission line and also  emission from adjacent lines.
\ion{O}{1} 1305 \AA\ is weak because of wind
absorption due to \ion{O}{1} 1306 \AA.  Similarly, \ion{C}{2} 1334.5 \AA\
is barely detected because of absorption in \ion{C}{2} 1335.7, and two
H$_2$ lines are weak because of absorption in \ion{C}{2} 1334.5.  Line
wavelengths at the radial velocity of RU Lupi are indicated by the vertical
dotted lines.  The narrow absorption feature near line center is produced
by the ISM.}
\end{figure}

\begin{figure}
\plotone{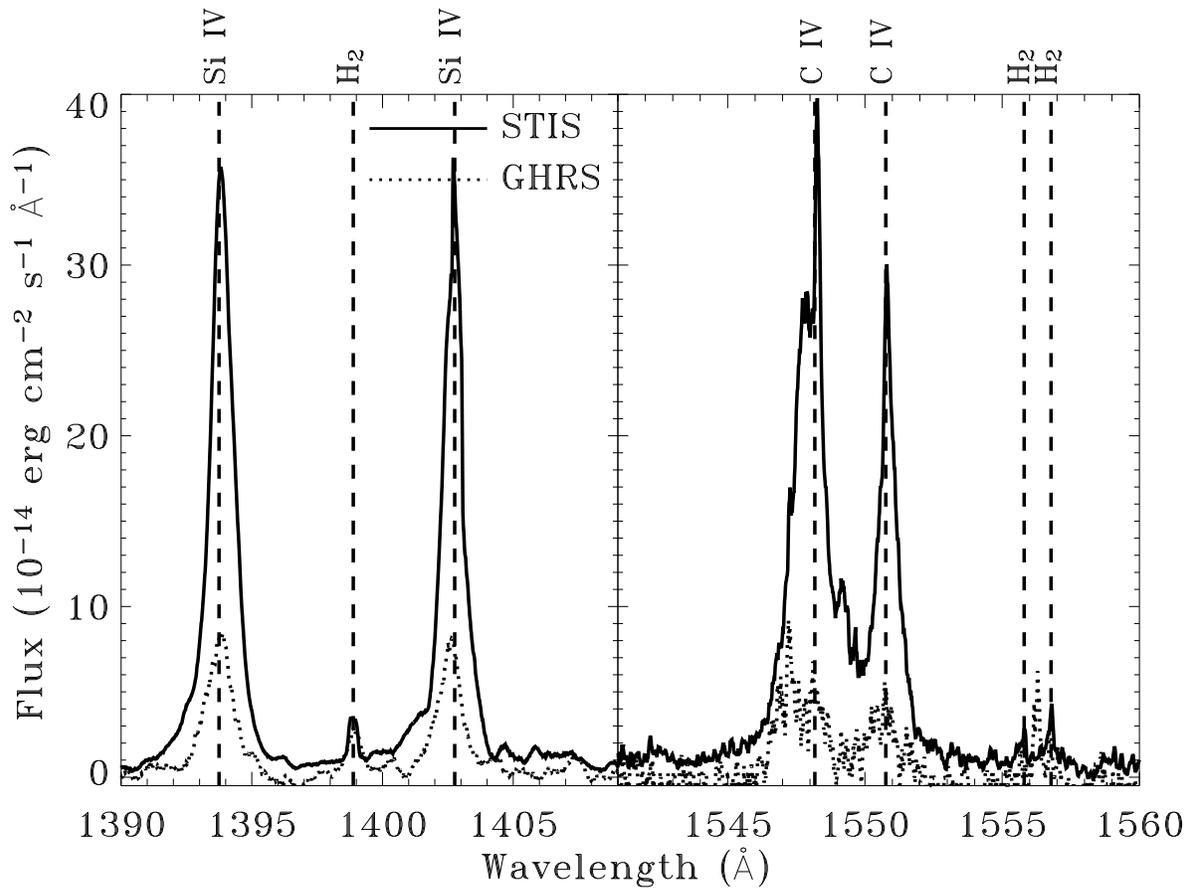}
\caption{The STIS spectrum of RU Lupi has a stronger FUV
continuum and stronger emission lines than the previously observed GHRS spectrum \citep{Ard02a}.}
\end{figure}

\clearpage
% Tables

\begin{table}
\caption{OBSERVATIONS OF RU LUPI}
\begin{tabular}{cccccc}
\hline
 Date & Instrument & Exposure (s) & Bandpass (\AA) & Aperture & Spectral Res. ($\lambda/\Delta\lambda$) \\
\hline
 7/12/00$^a$  & \HST/STIS E140M & 12530 & 1150--1700 & $0\farcs2\times 0\farcs06$ &45000\\
 7/12/00$^a$  & \HST/STIS G430L & 120 & 2900--5700 & $52\farcs\times 0\farcs$1& 800\\
 9/28/01$^b$ & \FUSE\ &  24517 & 900--1187 &
$30^{\prime\prime}\times30^{\prime\prime}$ & 15000\\
\hline
\multicolumn{2}{l}{$^a$JD 2451738} & \multicolumn{2}{l}{$^b$JD 2452180}\\
\end{tabular}
\end{table}

\begin{table}
\caption{{\it HST}/STIS OBSERVATIONS OF V471 TAU}
\begin{tabular}{cccccccc}
\hline
 Dates &  Orbits & Exposure & Aperture & $F_{1400}^a$ & Program & Obs. Numbers$^b$ & P.I.\\
\hline
3/13--3/23/98$^c$  & 6 & 14350 & $0\farcs2\times 0\farcs06$ &  4.01 & 7735 & 24906--24919 & F. Walter\\
8/24/00--8/27/00$^d$ &  4 & 7848 &  $0\farcs2\times 0\farcs2$& 4.46  & 8305 & 73220--73229 & H. Bond\\
1/25/02$^e$ &  4 & 10336 & $0\farcs2\times 0\farcs06$ & 4.01 & 9283 & 97310--97316 & F. Walter\\
\hline
\multicolumn{5}{l}{$^a$Peak flux from 1250--1600 \AA, in $10^{-10}$ erg
cm$^{-2}$ s$^{-1}$} &
\multicolumn{3}{l}{$^b$STIS observation numbers}\\ 
\multicolumn{3}{l}{$^c$JD 2450885--2450896} & \multicolumn{3}{l}{$^d$JD 2451781--2451784} & \multicolumn{2}{l}{$^e$JD 2452299}  \\
\end{tabular}
\end{table}

\begin{table}
\caption{Properties of Interstellar Absorption Lines$^a$}
\begin{tabular}{cccc}
\hline
ID & $\lambda_{obs}$ (\AA) & $v$ (\kms) & $W_\lambda$ (\AA)\\
\hline
%\ion{H}{1} & &&\\
\ion{Si}{3} & 1206.447  &  -11.2  & $0.089\pm0.007$\\
\ion{O}{1}    & 1302.090  &  -16.2  & $0.186\pm0.008$\\
\ion{C}{2}   & 1335.662  &  -8.3   & $0.098\pm0.005$\\
\ion{Si}{2}  & 1526.618  &  -15.1  & $0.172\pm0.009$\\
\ion{Fe}{2}  & 1608.367  &  -13.7  & $0.107\pm0.010$\\
\ion{Al}{2}  & 1670.672  &  -18.7  & $0.173\pm0.008$\\
\hline
\multicolumn{4}{l}{$^a$Measured by assuming Gaussian emission lines.}\\
\end{tabular}
\end{table}

\begin{table}
\caption{SPECTRA OF THE ACCRETION CONTINUUM}
\begin{tabular}{ccccc}
\hline
Bandpass & Instrument  & Scaling Factor$^a$ & S/N$^b$ & Date\\
\hline
1185--1700 & \HST/STIS E140M & 1.0 & 20 & 07/12/00\\
1700--1839 & \IUE\ SWP-LO & 0.75   & 60 & many$^c$\\
1839--1930 & \HST/STIS G230M & 1.1 & 35 & 08/24/92\\
1930--2302 &   \IUE\  SWP-LO& 0.75 & 14 & many$^c$\\
2302--2350 &  \HST/GHRS G270M & 1.1& 40 & 08/24/92\\
2350--2775 &  \IUE\  SWP-LO & 0.75 & 70 & many$^c$\\
2775--2822 & \HST/GHRS G270M & 1.1 & 120  & 08/24/92\\
2900--5700 & \HST/STIS G430L & 1.0 & 100 & 07/12/00\\
\hline
\multicolumn{5}{l}{$^a$Scaling factor to normalize flux to the {\it HST} fluxes.}\\
\multicolumn{5}{l}{$^b$Approximate S/N in a 5 \AA\ bin.}\\
\multicolumn{5}{l}{$^c$Multiple \IUE\ spectra were coadded by \citep*{Val03}
and}\\
\multicolumn{5}{l}{are available at http://archive.stsci.edu/prepds/iuepms/datalist.html.}
\end{tabular}
\end{table}

{\small
\begin{table}
\caption{Properties of emission lines in the \STIS\ spectrum$^a$}
\begin{tabular}{|ccccc|ccccc|}
\hline
ID & $\lambda_{obs}$  & $v$ & FWHM & Flux$^b$ & ID & $\lambda_{obs}$ & $v$ & FWHM  & Flux$^b$\\
 & \AA & \kms & \kms & & & \AA & \kms & \kms &\\
\hline
\ion{C}{3} &  1175.164 & - &  327 &   57.1 & \ion{Fe}{2} &  1360.112 & -12 &  128 &   9.5 \\
\ion{C}{3} &  1176.563 & -  &  103 &   14.4 & \ion{Fe}{2} &  1366.344 & -9 &  197 &   12.0 \\ 
\ion{Si}{2} &  1190.662 & 64 &  134 &    3.5 &   \ion{Fe}{2} &  1368.653 & -32 &  102 &    3.2\\
\ion{Si}{2} &  1194.462 & -8  &   75 &    3.7 &    \ion{Fe}{2}? &  1379.474 & 3 &  210 &    8.0\\
\ion{Si}{3} (1)$^d$ &  1206.149 & -85 &  105 &   24.8  & \ion{Si}{4} (1)$^d$&  1393.860 & 25 &  310 &  240\\
\ion{Si}{3} (2)$^d$ &  1206.495 & 1 &  362 &  214 &    \ion{Si}{4} (2)$^d$ &  1393.861 & 25 &  158 &  150\\
\ion{H}{1}$^c$ &  1217.500 & - &   -  & 75  &  \ion{Si}{4} (1)$^d$&  1402.636 & -27 &  113 &   78\\
\ion{N}{5} (1)$^d$ &  1238.151 & -160 &  368 &   28.0  &\ion{Si}{4} (2)$^d$ &  1402.704 & -12 &  275 &  225\\
\ion{N}{5} (2)$^d$ &  1238.828 & 4 &  135 &   20.3 &   \ion{P}{1}?&  1411.783 & 12 &  111 &    5.7\\
?  &  1252.154 &  - &   81 &    3.0 &  \ion{Fe}{2}? &  1414.920 & 10 &  172 &    4.5\\
\ion{Si}{2}$^c$ &  1265.000 & - &- & 11.3   &\ion{Cr}{2} &  1428.624 &-7  &   54 &    2.0\\
\ion{C}{1}? &  1280.956 & 28 &   77 &    1.5  &   \ion{Fe}{2}&  1464.918 & -11  &  118 &    5.3\\
\ion{Fe}{2} &  1288.641 & -17 &  228 &    5.8  &   \ion{S}{1} &  1471.847 & 5  &  115 &   13.7\\
\ion{Si}{3} &  1294.451 &  -20 &  127 &   10.1  &  \ion{C}{1}?&  1481.645 & -22 &  211 &    7.5\\
\ion{S}{1} &  1295.610 & -8 &  100  & 22.3     &   \ion{S}{1}?&  1483.426 & 21  &  213 &   12.2\\
\ion{S}{1} &  1296.056 &  -25 &  260  &  46.9   &  \ion{Fe}{2}&  1506.572 & 10 &  127 &    8.7\\
\ion{Si}{3} &  1298.992 & 13 &  203 &   22.8 &  \ion{Si}{2}$^c$&  1526.000 & - &    - & 14.2 \\
\ion{O}{1}$^c$ &  1302.000 & -  &- & 91   &   \ion{Si}{2}$^c$&  1533.000 & - &    - & 20 \\
\ion{O}{1}$^c$ &  1304.762 & -20 &   47 &   33.3 &   \ion{C}{4} (1)$^d$&  1547.900 & -54 & 202  &  173\\
\ion{O}{1}$^c$ &  1306.000 & - & - & 137  & \ion{C}{4} (2)$^d$&  1548.092 & -16  & 435  &   268\\
\ion{Si}{2}$^c$ &  1309.261 & -1 &  257 &   59 & \ion{C}{4} (3)$^d$&  1548.287 & 21  & 35 &  31.5\\
\ion{P}{2} &  1309.870 & 1  &  106 &   18.8 & \ion{C}{4} (1)$^d$&  1550.807 & 9 &229  & 192\\
\ion{Si}{2}$^e$ &  1310.747 & 12 &  232 &   16.0 & \ion{C}{4} (2)$^d$&  1550.839 &15 & 65  &  40\\
\ion{Fe}{2}$^f$ &  1312.630 & 34&  238 &    5.6 & \ion{Fe}{2}$^c$&  1558.928 & -28 &  161 &    7.2\\
\ion{S}{1} &  1323.355 & -34 &  141 &    4.2 & \ion{Fe}{2}$^c$&  1563.543 & -45  &   91 &    6.1\\
\ion{S}{1} &  1326.572 & -14 &   92 &    1.5 & \ion{Fe}{2}$^c$&  1566.486 & 35 &  211 &   13.3\\
\ion{C}{1} &  1329.518 & -11 &  189 &    5.2 & \ion{Fe}{2}&  1631.005 &-17  &  144 &   12.9\\
\ion{Fe}{2}? &  1330.985 & 9 &  146 &    4.8 &\ion{He}{2} (1)$^d$&  1640.455 & -1 &  256 &   35\\
\ion{C}{2}$^c$ &  1334.000 & -  &  - & 6.3 & \ion{He}{2} (2)$^d$ &  1640.499 & 7 &   70 &   28\\
\ion{C}{2}$^c$ &  1335.000 & - &  - & 112  &  \ion{He}{2} (3)$^d$ &  1641.284 & 150 &  124 &   12.7\\
\ion{Cr}{2} &  1347.136 & 24  &   83 &    2.9 &  \ion{O}{3}]&  1666.228 & 16 &  237 &   47\\
\ion{Cl}{1} &  1351.649 & 0 &  135 &   27.2 &  \ion{Al}{2}$^c$ &  1670.000 & - &  - &  15\\
\ion{O}{1} &  1355.615 & 6 &  157 &   35.5 &  \ion{Cr}{2}? &  1697.215 & -34 &  184 &   15.5\\
\ion{C}{1}? &  1357.240 & 25  &   93 &    2.0 & \ion{Fe}{2}$^c$&  1702.000 & - &   - &   23.6\\
\ion{O}{1} &  1358.438 & -14 &   88 &    5.5 & &&&&\\
\hline
\multicolumn{6}{l}{$^a$Lines fit with single Gaussian, except where noted.}
& \multicolumn{3}{l}{$^b10^{-15}$ erg cm$^{-2}$ s$^{-1}$}\\
\multicolumn{6}{l}{$^c$Wind absorption corrupts the emission line profile.}
&
\multicolumn{3}{l}{$^d$Fit with multiple Gaussians}\\
\multicolumn{6}{l}{$^e$Possibly blended with \ion{Si}{1}, \ion{Si}{2} emission.}
&
\multicolumn{3}{l}{labeled (1), (2), or (3)}\\
\multicolumn{9}{l}{$^f$Possibly blended with \ion{S}{1}}\\
\end{tabular}
\end{table}

{\small
\begin{table}
\caption{Atomic and unidentified Emission Lines in the \FUSE\ spectrum$^a$}
\begin{tabular}{ccccc}
\hline
ID & $\lambda_{obs}$ (\AA) & $v$ (\kms) & FWHM (km s$^{-1}$) & Flux$^b$\\
\hline
\ion{C}{3} (1)$^{c,d}$ &    976.154 & -263 &  106 &    3.7\\
\ion{C}{3} (2)$^{c,d}$ &    976.679 & -102 &   95 &    10.5\\
 ? &    997.313 & - &   36 &    0.8\\
\ion{O}{6} &   1031.251 & -190 &  359 &    30.4\\
\ion{O}{6}$^e$ &   1036.861 & -215 &   52 &   2.8\\
?  &   1048.151 & - &   67 &    2.3\\
?  &   1074.385 & - &   42 &    1.3 \\
\ion{N}{1}$^c$ &   1134.284 & -33 &  128 &  5.4\\
\ion{N}{1}$^c$ &   1134.974 & 0 &   64 &   3.3\\
\ion{N}{1}?&   1168.514 & -4 &   79 &    1.5\\
\hline
\multicolumn{5}{l}{$^a$Lines fit with single Gaussian.}\\
\multicolumn{5}{l}{$^b10^{-15}$ erg cm$^{-2}$ s$^{-1}$}\\
\multicolumn{5}{l}{$^c$Wind absorption corrupts emission profile.}\\
\multicolumn{5}{l}{$^d$Fit with multiple Gaussians.}\\
\multicolumn{5}{l}{$^e$Corrupted by H$_2$ absorption.}\\
\end{tabular}
\end{table}

\begin{table}
\caption{Pumped atomic lines in the STIS spectrum}
\begin{tabular}{ccccc}
\hline
ID & $\lambda_{obs}$ (\AA) & ID$_{pump}$ & $\lambda_{pump}$ (\AA)\\
\hline
\ion{Fe}{2} & 1288.641 & \ion{H}{1} & 1213.738 \\
\ion{S}{1} & 1295.610 & \ion{O}{1} & 1302.336 \\
\ion{S}{1} & 1296.056 & \ion{O}{1} & 1302.862\\
\ion{Si}{2} & 1309.261 & \ion{O}{1} & 1304.370\\
\ion{P}{2} & 1309.870 & \ion{O}{1} & 1301.874, 1304.675\\
\ion{P}{2} & 1310.747 & \ion{O}{1} & 1305.497\\
\ion{Fe}{2} & 1312.630 & \ion{H}{1} & 1215.873\\
\ion{Fe}{2}? & 1330.985 & \ion{O}{1} & 1304.436\\
\ion{Cr}{2} & 1347.149 & \ion{H}{1} & 1215.765\\ 
\ion{Cl}{1} & 1351.651 & \ion{C}{2} & 1335.726\\
\ion{Fe}{2} & 1360.117 & \ion{H}{1} & 1215.503\\
\ion{Fe}{2} & 1366.345 & \ion{H}{1} & 1214.735\\ 
\ion{Fe}{2} & 1368.649 & \ion{H}{1} & 1214.735\\
\ion{Fe}{2}? & 1379.474 & \ion{Si}{4}? & 1403.255\\
\ion{P}{1}? & 1411.783 & \ion{H}{1} & 1216.608\\
\ion{Fe}{2}? & 1414.920 & \ion{C}{3} & 1175.147\\
\ion{Cr}{2} & 1428.624 & \ion{H}{1} & 1217.387\\
\ion{Fe}{2} & 1464.918 & \ion{H}{1} & 1213.738 \\
\ion{S}{1} & 1471.840 & \ion{O}{1} & 1302.336 \\
\ion{Fe}{2} & 1506.544 & \ion{C}{2} & 1335.409\\
\ion{Cr}{2}? & 1697.215 & \ion{H}{1} & 1304.381\\
\hline
\end{tabular}
\end{table}

\begin{table}
\caption{Identification of H$_2$ lines in the \STIS\ spectrum}
\begin{tabular}{|cccc|cccc|ccc|}
\hline
ID & $\lambda_{obs}$ (\AA) & Flux$^a$ && ID & $\lambda_{obs}$ (\AA) & Flux$^a$&& ID & $\lambda_{obs}$ (\AA) & Flux$^a$ \\
\hline
1-2 R(3) &   1202.396 &     3.1 && 
1-4 R(6) &   1327.506 &     1.7 && 
1-7 R(3) &   1489.497 &    10.6 \\
1-5 R(3)$^b$ &   1208.787 &     1.8 && 
1-4 P(5) &   1329.047 &    -1.0 && 
4-8 P(8) &   1494.917 &     1.2 \\
0-2 P(3) &   1225.469 &     2.1 && 
0-4 P(2) &   1338.512 &    13.0 && 
1-7 R(6) &   1500.364 &     8.6 \\
2-2 R(11) &   1237.511 &     1.2 && 
0-4 P(3) &   1342.192 &    12.9 && 
1-7 P(5) &   1504.679 &    18.5 \\
1-2 P(8) &   1237.827 &     4.5 && 
4-6 R(3) &   1359.031 &     1.4 && 
4-9 R(3) &   1513.489 &     1.7 \\
4-4 R(3) &   1253.625 &     1.5 && 
4-6 R(6) &   1371.055 &     0.7 && 
3-7 R(15) &   1513.916 &     1.5 \\
3-1 P(17) &   1254.074 &     1.7 && 
4-4 R(17) &   1372.014 &     0.9 && 
blend$^e$ &   1516.122 &     4.6 \\
1-3 R(3) &   1257.780 &     6.4 && 
4-6 P(5) &   1372.671 &     1.3 && 
0-7 P(2) &   1521.529 &     1.8 \\
3-2 R(15) &   1265.121 &     0.7 && 
2-4 P(13) &   1379.913 &     1.1 && 
1-7 P(8) &   1524.586 &     8.8 \\
4-4 R(6) &   1266.515 &     0.6 && 
1-5 P(5) &   1387.299 &     3.4 && 
0-7 P(3) &   1525.092 &     1.5 \\
4-4 P(5)$^c$ &   1266.823 &     1.7 && 
0-5 P(1) &   1396.157 &     0.9 && 
2-8 R(11) &   1555.810 &     3.2 \\
3-1 P(18) &   1268.806 &     0.5 && 
0-5 P(2) &   1398.883 &    10.8 && 
1-8 R(6) &   1556.796 &     7.5 \\
3-4 P(1) &   1270.527 &     0.6 && 
2-5 R(11) &   1399.151 &     1.8 && 
1-8 P(5) &   1562.317 &     6.9 \\
1-3 R(6) &   1270.967 &     3.1 && 
0-5 P(4) &   1407.154 &     3.2 && 
4-10 P(5) &   1572.548 &     1.6 \\
2-2 P(13) &   1271.128 &     2.4 && 
blend$^d$ &   1408.807 &     1.8 && 
1-8 P(8) &   1580.573 &     4.8 \\
1-3 P(5) &   1271.869 &     5.1 && 
3-5 R(15) &   1418.169 &     1.1 && 
0-8 P(2) &   1582.350 &     3.1 \\
4-2 R(17) &   1273.968 &     1.7 && 
1-6 R(3) &   1430.945 &     9.6 && 
2-8 P(13) &   1588.709 &     6.7 \\
0-3 R(0) &   1274.495 &     6.4 && 
2-5 P(13) &   1434.439 &     2.2 && 
3-10 P(1) &   1591.263 &     1.3 \\
0-3 R(1) &   1274.855 &     4.5 && 
1-6 R(6) &   1442.794 &     5.8 && 
3-9 R(15) &   1593.191 &     3.9 \\
0-3 R(2) &   1276.222 &     1.0 && 
1-6 P(5) &   1446.051 &    14.2 && 
2-9 R(11) &   1602.176 &     4.9 \\
3-2 R(16) &   1277.958 &     0.6 && 
2-6 R(11) &   1453.024 &     3.4 && 
4-11 R(3) &   1602.543 &     3.0 \\
0-3 P(2) &   1279.407 &     5.9 && 
0-6 R(0) &   1454.761 &     3.7 && 
1-9 R(3) &   1603.177 &     2.5 \\
0-3 P(3) &   1283.051 &     6.1 && 
0-6 R(1) &   1454.894 &     5.1 && 
4-11 R(6) &   1606.202 &     2.1 \\
0-3 P(4) &   1287.632 &     0.4 && 
0-6 R(2) &   1455.847 &     1.0 && 
1-9 R(6) &   1610.884 &     2.5 \\
4-4 P(8) &   1287.886 &     0.6 && 
0-6 P(1) &   1457.348 &     1.1 && 
4-11 P(5) &   1613.645 &     4.8 \\
2-3 R(11) &   1290.842 &     0.8 && 
0-6 P(2) &   1460.095 &     7.6 && 
1-9 P(5) &   1617.817 &     2.3 \\
1-3 P(8) &   1293.814 &     4.4 && 
0-6 P(3) &   1463.712 &    11.6 && 
3-9 P(17) &   1622.084 &     1.5 \\
1-4 R(3) &   1314.570 &     2.6 && 
4-6 R(17) &   1465.096 &     2.1 && 
4-11 P(8) &   1622.821 &     3.8 \\
3-2 P(18) &   1320.133 &     0.7 && 
1-6 P(8) &   1467.014 &     6.2 && 
2-9 P(13) &   1632.458 &     3.1 \\
3-5 P(1) &   1324.966 &     0.8 && 
4-8 P(5) &   1476.996 &     1.0 && 
3-11 P(1) &   1636.275 &     3.9 \\
\hline
\multicolumn{4}{l}{$^a10^{-15}$ erg cm$^{-2}$ s$^{-1}$} & \multicolumn{7}{l}{$^b$Werner-band (C-X), pumped by \ion{O}{6} at 1031.862 \AA}\\
 \multicolumn{5}{l}{$^c$Also includes emission from 4-1 P(19)} &
 \multicolumn{6}{l}{$^d$Blend of 1-5 P(8) and 3-4 P(17)}\\
 \multicolumn{7}{l}{$^e$Blend of 0-1 R(0) and 0-1 R(1)}\\
 \end{tabular}
\end{table}

{\small
\begin{table}
\caption{Lyman and Werner-band H$_2$ lines in the \FUSE\ spectrum}
\begin{tabular}{cccccccccc}
\hline
ID$^a$ & $\lambda_{obs}$ (\AA) & $v$ (\kms) & FWHM (\kms) & Flux$^b$\\
\hline
C-X 2-2 P(5) & 1058.255 & -22 & 31 & 1.3 &\\
B-X 1-0 P(5) &   1109.273 & -8&   36 &    0.4 &\\
C-X 1-3 Q(3) &   1119.017 & -14 &   71 &    0.7 &\\
C-X 0-2 Q(10)+ C-X 1-3 P(5) & 1127.096 & - & 61 & 1.5\\
B-X 1-1 R(3) &   1148.632 & -16 &   30 &    1.4 &\\
C-X 1-3 P(11) &   1154.850 & -14 &   36 &    1.5 &\\
B-X 1-1 P(5)+ B-X 1-1 R(6)$^c$&   1161.774 & - &   60 &   3.4 &\\
B-X 0-1 P(2) &   1166.176 & -18 &   34 &    0.9 &\\
B-X 0-1 P(3) & 1169.687 & -15 & 38 & 0.8 & \\ 
C-X 0-3 Q(10) + C-X 1-4 P(5) &   1171.916 & - &   47 &    3.4 &\\
blend$^d$ & 1174.396 & - &  122 &    4.9\\
C-X 2-6 R(3) & 1174.786 & -19& 25 & 1.6 & \\
C-X 1-3 Q(16) & 1177.208 & -14& 21 &0.6 & \\
\hline
\multicolumn{5}{l}{$^a$B-X refers to Lyman-band, C-X refers to Werner-band}\\
\multicolumn{4}{l}{$^b10^{-15}$ erg cm$^{-2}$ s$^{-1}$}\\
\multicolumn{5}{l}{$^c$B-X 0-1 R(0) and B-X 0-1 R(1) may also contribute}\\
\multicolumn{5}{l}{$^c$C-X 1-5 Q(7), C-X 1-4 R(9), C-X 2-5 R(0), C-X 2-5
R(1), C-X 2-5 R(2)}\\
\end{tabular}
\end{table}

\begin{table}
\caption{Properties of Wind Absorption Lines$^a$}
\begin{tabular}{cccccc}
ID & $\lambda_{obs}$ (\AA) &  $v_c^b$ (\kms) & $ v_{max}^c$ (\kms) & \Elo$^d$ (cm$^{-1}$) & $W_\lambda$ (\AA)\\
\hline
\ion{Si}{2}  & 1189.75 & -170 & -290 &  0   & $1.20\pm0.07$\\
\ion{N}{1}    & 1199.83 &  -   &  -   &  0   & $1.06\pm0.11^e$\\
\ion{Si}{3} & 1206.75 & -180 & -220 &  0   & $0.20\pm0.02$ \\
\ion{N}{1}    & 1242.74 & -110 & -190 &19224 & $0.86\pm0.04$\\
\ion{Si}{2}  & 1265    & -150 &  -   & 287  & $1.15\pm0.15^f$\\
\ion{O}{1}    & 1301.43 & -170 & -290 &  0   & $0.94\pm0.01$ \\
\ion{Si}{2}  & 1303.57 & -180 & -260 &  0   & $0.61\pm0.10^g$\\
\ion{O}{1}    & 1304.30 &  -130 & -220 & 158  & $0.72\pm0.10^g$\\
\ion{O}{1}    & 1305.30 & -170 & -280 & 227  & $0.88\pm0.01$ \\
\ion{Si}{2}  & 1308.55 & -170 & -270 & 287  &  $0.78\pm0.05$\\
\ion{C}{2}   & 1334.5  & -140 & -230 &  0   & $0.98\pm0.25$\\
\ion{C}{2}   & 1334.93 & -170 & -300 & 63   & $1.08\pm0.01$ \\
\ion{Cl}{1}   & 1346.35 & -200 & -320 &  0   &  $0.38\pm0.10$\\
\ion{S}{1}    & 1424.60 & -120 & -210 &  0   & $0.27\pm0.04$\\
\ion{N}{1}    & 1492.02 & -120 & -210 & 19224& $0.77\pm0.05$\\
\ion{N}{1}    & 1493.75 & -190 & -310 & 19224& $0.56\pm0.03$\\
\ion{Si}{2}  & 1525.84 & -170 & -280 &  0   & $0.96\pm0.05$\\
\ion{Si}{2}  & 1532.55 & -170 & -280 & 287  & $0.97\pm0.05$\\
\ion{Fe}{2} & 1558.18 & -170 & -270 & 1872 & $0.60\pm0.09$\\
\ion{Fe}{2} & 1562.93 & -160 & -260 & 2430 & $0.50\pm0.05$\\
\ion{Fe}{2} & 1565.95 &  -70 & -140 & 1872 & $0.40\pm0.10$\\
\ion{Fe}{2} & 1574.23 & -150 & -200 & 1872 & $0.23\pm0.05$\\
\ion{Fe}{2} & 1579 59 & -190 & -290 & 2430 & $0.53\pm0.04$\\
\ion{Fe}{2} & 1607.55 & -170 & -270 &  0   & $0.66\pm0.08$\\
\ion{Fe}{2} & 1612.79 & -190 & -270 & 1872 & $0.55\pm0.09$\\
\ion{Fe}{2} & 1620.73 & -170 & -230 & 385  & $0.34\pm0.04$\\
\ion{Fe}{2} & 1624.64 & -160 & -260 & 2430 & $0.47\pm0.06$\\
\ion{Fe}{2} & 1628.15 & -180 & -250 & 667  & $0.28\pm0.07$\\
\ion{Al}{2}  & 1669.85 & -170 & -270 & 0    & $1.02\pm0.06$\\
\ion{Fe}{2} & 1701.22 & -140 & -200 & 1872 & $0.53\pm0.05$\\
\hline
\multicolumn{6}{l}{$^a$Measured assuming an emission line symmetric about the radial velocity of the star}\\
\multicolumn{6}{l}{or as absorption against the continuum.}\\
\multicolumn{3}{l}{$^b$Central velocity of absorption.} &
\multicolumn{3}{l}{$^c$Maximum velocity of absorption.}\\
\multicolumn{2}{l}{$^d$Energy of lower level.} &
\multicolumn{1}{l}{$^e$Multiplet.} &
 \multicolumn{2}{l}{$^f$Uncertain blue edge} &  \multicolumn{1}{l}{$^g$Blend}\\
\end{tabular}
\end{table}

\end{document}